\documentclass{article}

\usepackage{amsmath}
\usepackage[ruled]{algorithm2e}
\usepackage{url}

\usepackage{pdflscape} 

\usepackage{tikz}
\usetikzlibrary{shapes,arrows,arrows.meta, fit}
\usetikzlibrary{positioning,calc}
\usepackage{float, caption}

\usepackage[OT1]{fontenc}
\usepackage{amsthm,amsmath}
\usepackage{natbib}

\usepackage{xr}
\usepackage{mathrsfs, bm}
\usepackage{float}
\usepackage[margin=1.3in]{geometry}

\def\bm{\boldsymbol}
\def\bz{\bm z}
\def\by{\bm y}
\def\bx{\bm x}
\def\bs{\bm s}
\def\bA{\bm A}
\def\bSigma{\bm \Sigma}

\def\hTh{\hat{\Theta}}
\def\bbeta{\bm \beta}

\def\vect{\mbox{vec}}

\def\hTh{\boldsymbol{\Theta}}

\theoremstyle{plain}

\usepackage{lipsum}

\newcommand\blfootnote[1]{%
	\begingroup
	\renewcommand\thefootnote{}\footnote{#1}%
	\addtocounter{footnote}{-1}%
	\endgroup
}

\begin{document}
	
	\author{Joshua Lukemire, Yikai Wang, Amit Verma, and Ying Guo \\ Emory University}
	\title{HINT: A Hierarchical Independent Component Analysis Toolbox for Investigating Brain Functional Networks using Neuroimaging Data.}
	\date{}
	
	\maketitle
	
	\begin{abstract}
	
	{\bf{\underline{Background}}}\\
			Independent component analysis (ICA) is a popular tool for investigating brain organization in neuroscience research. In fMRI studies, an important goal is to study how brain networks are modulated by subjects' clinical and demographic variables. { Existing ICA methods and toolboxes don't incorporate subjects' covariates effects in ICA estimation of brain networks, which potentially leads to loss in accuracy and statistical power in detecting brain network differences between subjects' groups.}
			
		{\bf{\underline{New Method}}}	\\
			 We introduce a Matlab toolbox, {\bf HINT} ({\bf H}ierarchical {\bf IN}dependent component analysis {\bf T}oolbox), that provides a hierarchical covariate-adjusted ICA (hc-ICA) for modeling and testing covariate effects and generates model-based estimates of brain networks on both the population- and individual-level. HINT provides a user-friendly Matlab GUI that allows users to easily load images, specify covariate effects, monitor model estimation via an EM algorithm, specify hypothesis tests, and visualize results. HINT also has a command line interface which allows users to conveniently run and reproduce the analysis with a script.
			
		{\bf{\underline{Comparison to Existing Methods}}}	\\
		    HINT implements a new multi-level probabilistic ICA model for group ICA. It provides a statistically principled ICA modeling framework for investigating covariate effects on brain networks. HINT can also generate and visualize model-based network estimates for user- specified subject groups, which greatly facilitates group comparisons.
		    
		{\bf{\underline{Results}}} \\
		    We demonstrate the steps and functionality of HINT with an fMRI example data to estimate treatment effects on brain networks while controlling for other covariates. Results demonstrate estimated brain networks and model-based comparisons between the treatment and control groups. In comparisons using synthetic  fMRI data, HINT shows desirable statistical power in detecting group differences in networks especially in small sample sizes, while maintaining a low false positive rate. HINT also  demonstrates similar or increased accuracy in reconstructing both population- and individual-level source signal maps as compared to some state-of-the-art group ICA methods.
			
		{\bf{\underline{Conclusion}}}	\\
			HINT can provide a useful tool for both statistical and neuroscience researchers to evaluate and test differences in brain networks between subject groups.
		
	\end{abstract}
	
	
\section{Introduction}
\blfootnote{HINT can be downloaded from NITRC at: \url{https://www.nitrc.org/projects/hint}}
In recent years, there has been a growing interest in network based approaches to investigate brain organization and function. Under a network approach, observed brain signals represent a combination of signals generated from distinct brain functional networks. For example, in functional magnetic resonance imaging (fMRI), the observed blood-oxygen-level-dependent (BOLD) signal can be viewed as a combination of contributions from different brain networks. These brain functional networks can help reveal  the brain's functional organization structure as they have been shown to be present across a wide variety of subjects and across a range of different experimental conditions \citep{smith2009, smith2013, kemmer2015network}. Consequently, studying these brain functional networks has become a topic of great interest \citep{ma2007detecting,biswal2010toward,wang2016efficient,kemmer2018evaluating}. Currently, one of the most popular brain network estimation tools is independent component analysis (ICA). As a special case of blind source separation, ICA identifies brain functional networks by separating observed imaging data, mostly fMRI signals, into linear combinations of latent source signals that are assumed to be statistically independent and non-Gaussian.

The earliest usage of ICA for studying brain networks applied Spatial ICA to single subject fMRI data. Spatial ICA  \citep{mckeown1997} separates the observed data into independent spatial maps and corresponding time courses and is the most popular technique for brain network estimation due to its ease of interpretation and the high spatial resolution of fMRI data. For fMRI data with a large number of measurements across time, it is also possible to perform temporal ICA \citep{smith2012}, which separates the observed data into temporally coherent maps.

While methods for single subject ICA are quite well established, applying ICA to multiple subjects or groups is not straightforward and requires extension of the original ICA. Existing group level ICA techniques are frequently based on temporal-concatenation group ICA (TC-GICA), which assumes that brain functional networks have the same spatial pattern across different subjects. Under a TC-GICA approach, the subject level data are stacked in the time domain and the spatial source signals (ICs) are extracted. The subject level ICs are then recovered using techniques such as back-reconstruction \citep{calhoun2001, erhardt2011comparison} or dual regression \citep{beckmann2009}. Inference on covariate effects under these frameworks requires secondary hypothesis testing or regression analysis.

Recently, Shi and Guo \citep{shi2016} proposed a hierarchical-covariate adjusted spatial ICA (hc-ICA) framework for estimating underlying brain networks adjusted by subject-specific covariate effects. This novel technique is fundamentally different from other group ICA methods because it directly incorporates covariate effects in the ICA decomposition. In this work we introduce the {\bf H}ierarchical {\bf IN}dependent component analysis {\bf T}oolbox (HINT), a Matlab toolbox aimed at implementing these powerful hierarchical ICA techniques in a user-friendly platform allowing researchers to easily conduct analyses under the hierarchical ICA framework. HINT is released as an open source package under the MIT license (https://opensource.org/licenses/MIT).  By implementing the advanced hc-ICA model, HINT aims to improve the accuracy in estimation of brain functional networks by modeling covariate effects to account for between-subject heterogeneity in networks, which are not modelled in ICA decomposition of the current TC-GICA methods. Simulation studies have shown that this model based approach can result in improved power to detect differences in sub-populations \citep{shi2016, wang2019hierarchical}, especially in small size scenarios. Throughout this paper, we use the term ``sub-populations" to refer to sub-groups defined by user-specified clinical and demographic characteristics, such as treatment and control groups.

The hc-ICA implemented in HINT has several appealing features:  it can provide model-based estimates and prediction of sub-population brain networks and it offers a formal statistical framework for estimating and testing covariate effects. Such a statistically principled framework can potentially lead to improved accuracy and increased power for detecting brain network differences. For example, we \citep{shi2016} showed that hc-ICA identified significant differences in the visual network between subjects in groups with and without post traumatic stress disorder whereas an existing probabilistic TC-GICA method did not. The findings from hc-ICA coincided with the results reported in previous fMRI studies which showed enhanced activities in visual cortex among PTSD subjects. As another example, in our recent work \cite{wang2019hierarchical}, we applied a longitudinal version of hc-ICA to investigate the differences between Alzheimer's disease patients and control subjects. In comparison, we applied a probabilistic TC-GICA \citep{beckmann2005} with dual regression method. The hc-ICA framework identified more significant differences across major resting state networks. In particular, we identified significant differences in the default mode network between Alzheimer's disease patients and normal controls which coincided with previous findings in the literature, while the probabilistic TC-GICA wasn't able to detect such differences from the data. In both data applications, the hc-ICA results have been shown to be consistent across different sets of initial values and robust to leave-one-out validation analyses.

At this time, the toolbox implements the hc-ICA technique of \cite{shi2016}, with further extensions of this approach,  including the longitudinal version of hc-ICA \cite{wang2019hierarchical}, in active development. The intent of this toolbox is to provide neuroimaging researchers with an easy-to-use tool for utilizing hierarchical ICA techniques on fMRI data, allowing them to estimate the brain networks of interest, test hypotheses about covariate effects, and generate model-based estimates of the brain networks on both the study population- and individual-level. The toolbox features several interactive visualization windows that allow users to flexibly specify the brain networks and subject sub-populations for results visualization. HINT also features a computationally optimized EM algorithm with significant improvements over the original EM in the hc-ICA paper \citep{shi2016}. This new EM algorithm in HINT provides much more efficient estimation of the hc-ICA model, particularly on data sets with a large number of subjects.

The HINT Matlab GUI allows the user to input their imaging data in The Neuroimaging Informatics Technology Initiative (NIFTI) file format. HINT then performs the preprocessing steps prior to hc-ICA, obtains initial guesses for the hc-ICA model parameters, and, upon user request, removes ICs that are not of interest, such as those corresponding to motion artifacts, from the subsequent hc-ICA modeling. Model estimation within HINT is carried out via an Expectation Maximization (EM) algorithm as described in Section \ref{emAlgs}. HINT can perform hypothesis testing on the covariate effects and linear combinations of the covariate effects using a voxel-specific approximate inference procedure \citep{shi2016}.  After obtaining parameter estimates, the toolbox provides  display windows enabling the user to visualize brain network maps in several configurations. Specifically, the user can visualize the model-based estimates of the brain functional networks for sub-populations of interest by specifying their corresponding covariate patterns. The user can also visualize the model-based estimates of subject-specific brain networks to obtain individual-level network information.  Furthermore, users can use the covariate effect display window to identify specific regions of the brain that show significant differences between groups, as well as perform model-based hypothesis testing of contrasts of those covariate effects.

In addition to the GUI interface, the HINT toolbox also includes a command line interface for the hc-ICA analysis allowing users to easily reproduce the analysis with a script or run HINT on a high performance computing cluster.

This paper proceeds as follows. First, Section 2 provides a review of the hc-ICA model, a description of the EM algorithm for estimating the model parameters, and an introduction to inference procedure for the covariate effects. Then, in Section 3, we provide a discussion of the toolbox. In Section 4, we carry out a simulation study using synthetic data in order to compare HINT to TC-GICA. Finally, in Section \ref{sec:summary} we discuss future directions for HINT and provide some concluding remarks. In the Appendix, we provide a detailed walkthrough of the toolbox features, explain how they interface with the statistical model, and introduce the various visualization windows.

\section{Methods}\label{sec:model}

In this section, we describe the details of the hc-ICA method. Specifically, we discuss the required preprocessing, the hc-ICA model, and hypothesis testing on the covariate effects. All aspects of this procedure are implemented in HINT.

\subsection{Preprocessing}\label{subsubsec::preproc}
As in standard ICA methods, prior to hc-ICA analysis, some preprocessing steps such as centering, dimension reduction and whitening of the observed data are performed by the toolbox to facilitate the subsequent ICA decomposition \citep{hyvarinen2000}. Let $\tilde{\boldsymbol{Y}}_i$ be the $T\times V$ fMRI data matrix for subject $i$ where $T$ is the number of fMRI scans and $V$ is the number of voxels in a 3D fMRI scan. In HINT, prior to hc-ICA, each  subject's observed fMRI images are preprocessed as follows \citep{shi2016},
\begin{equation}\label{preprocEqn}
\boldsymbol{Y}_i = ( \boldsymbol{\Lambda}_{i,q} - \sigma_{i,q}^2 \boldsymbol{I}_q )^{-1/2} \boldsymbol{U}_{i,q}^T \tilde{\boldsymbol{Y}}_i,
\end{equation}
where $q$ is the number of independent components, $\boldsymbol{\Lambda}_{i,q}$ contains the first $q$ eigenvalues and $\boldsymbol{U}_{i,q}$ contains the first $q$ eigenvectors as obtained by a singular value decomposition (SVD) of the original fMRI data $\tilde{\boldsymbol{Y}}_i$. The residual variance, $\sigma_{i,q}^2$ , is estimated by the average of the $T-q$ smallest eigenvalues that are not included in $\bm \Lambda_{i, q}$, representing the variability in $\widetilde{\bm Y}_i$ that is not accounted for by the first $q$ components. The number of independent components, $q$, is specified by the user. This can be determined based on theoretical methods such as the Laplace approximation method \citep{minka2001}, using the IC number suggested by existing group ICA toolboxes or based on established knowledge on ICA analysis of fMRI data.

\subsection{hc-ICA Model}

The hc-ICA framework decomposes multi-subject fMRI data using a two-level model, where the first level models each individual subject's data as a mixing of subject-specific independent components (ICs), and the second level models the subject-specific ICs as a function of population-level spatial source signals and the covariate effects. More specifically, at the first level, each subject's preprocessed data is decomposed into a linear mixture of subject-level ICs as:
\begin{equation}\label{firststage}
	\boldsymbol{y_i}(v) = \boldsymbol{A}_i \boldsymbol{s_i}(v) + \boldsymbol{e}_i (v),
\end{equation}
where $\boldsymbol{y_i}(v)$ is the column vector from $\boldsymbol{Y}_i$ that corresponds to voxel $v$, $\boldsymbol{s_i}(v) = [ s_{i1}(v), $ $s_{i2}(v),\ldots, s_{iq}(v) ]^T$ is a $q \times 1$ vector, and each $s_{il}(v) (l=1,\ldots,q)$ contains the spatial source signal of the $l$th IC at the $v$th voxel for subject $i$. $\boldsymbol{A}_i$ is the $q \times q$ orthogonal mixing matrix for subject $i$, which mixes the spatial source signals to generate the observed data, $\boldsymbol{e}_i (v)$ is a $q\times 1$ noise vector that represents the residual variations in the subject's data that are not explained by the $q$ extracted ICs. hc-ICA assumes that $ \boldsymbol{e}_i (v) \sim N( \boldsymbol{0}, \boldsymbol{E}_v )$. Moreover, since the prewhitening described in Section \ref{subsubsec::preproc} is performed to remove temporal correlations in the noise term and to standardize the variability across voxels, hc-ICA assumes that the noise covariance is identical across voxels and is isotropic \citep{beckmann2005, guo2011, shi2016},  i.e. $\bm E_v=\sigma^2_0 \bm I_q$.
		
The second-level of hc-ICA models the subject level spatial source signals as a combination of population-level spatial source signals, covariate effects, and subject-specific random variabilities:
\begin{equation}\label{secondstage}
	\boldsymbol{s_i}(v) =  \boldsymbol{s_0}(v) + \boldsymbol{\beta}(v)^T \boldsymbol{x}_i + \boldsymbol{\gamma}_i (v),
\end{equation}
where $\boldsymbol{s_0(v)} = [ s_{01}(v), s_{02}(v),\ldots, s_{0q}(v) ]^T$ contains population level source signals for each of the $q$ ICs. Covariate information is encoded in $\boldsymbol{x}_i = [ x_{i1}, x_{i2},\ldots, x_{ip} ]^T$, a $p \times 1$ vector with the covariate settings for subject $i$. The $q \times 1$ error vector $\boldsymbol{\gamma}_i (v)$ reflects the residual between-subject random variability after controlling for the covariate effects. hc-ICA assumes that $\boldsymbol{\gamma}_i (v) \sim N(\boldsymbol{0}, \boldsymbol{D})$, where $\boldsymbol{D} = \text{diag}( \nu_1^2, \ldots, \nu_q^2 )$. We allow the variances to be IC-specific to allow for different levels of between-subject random variability across various brain networks.

The population level spatial source signals, $\boldsymbol{s}_0(v)$, are modeled under a mixtures of Gaussians (MoG) approach \citep{shi2016,guo2011, guo2013,gao2017modeling,gao2018regularized,wang2019hierarchical}. That is, for IC $\ell, \; \ell = 1, \ldots, q$, we have

\begin{equation}\label{mog}
 s_{0\ell}(v) \sim \mbox{MoG}(\bm \pi_{\ell}, \bm \mu_{\ell}, \bm \sigma^2_{\ell}), \qquad v=1,...,V,
 \end{equation}
where $\bm \pi_{\ell}=[\pi_{\ell,1},...,\pi_{\ell,m}]'$ with $\sum_{j=1}^m \pi_{\ell, j}=1$, $\bm \mu_{\ell}=[\mu_{\ell,1},...,\mu_{\ell,m}]'$ and $ \bm \sigma^2_{\ell} = [\sigma^2_{\ell,1},...,\sigma^2_{\ell,m}]'$; $m$ is the number of Gaussian components in MoG. The probability density of $\mbox{MoG}(\bm \pi_{\ell}, \bm \mu_{\ell}, \bm \sigma^2_{\ell})$ is $\sum_{j=1}^m \pi_{\ell, j} g(s_{0\ell}(v); \mu_{\ell,j}, \sigma^2_{\ell, j})$ where $g(\cdot)$ is the pdf of the Gaussian distribution. In fMRI applications, mixtures of two to three Gaussian components are sufficient to capture the distribution of fMRI spatial signals, with the different Gaussian components representing the background fluctuation and the negative or positive fMRI BOLD effects respectively \citep{beckmann2004, guo2008}.

To facilitate derivations in models involving MoG, we define latent state variable  $\bz(v)=[z_{1}(v),...,z_q(v)]'$ at voxel $v$. For $\ell=1,...,q$,  $z_{\ell}(v)$ takes a value in $\{1,\ldots,m\}$ with probability $p[z_{\ell}(v)=j]=\pi_{\ell,j}$ for $j=1,..,m$. The latent state variable $z_{\ell}(v)$ represents the voxel $v$'s membership in the MoG of the $\ell$ IC, with $z_{\ell}(v)=j$ indicating that voxel $v$ comes from the $j$th Gaussian component in the MoG distribution in the $\ell$ network.

\subsection{EM Algorithm for model estimation} \label{emAlgs}
HINT employs the subspace-based EM algorithm in Shi and Guo (2016) to estimate the parameters in the hc-ICA model. Specifically, the model parameters are estimated using a unified maximum likelihood method via the EM algorithm that simultaneously estimates all parameters in the hc-ICA model. The detailed expression for the complete data log-likelihood function at each voxel $v$ is:
\begin{align}\label{likv}
\displaystyle l_v(\Theta ; \mathcal{Y},  \mathcal{X}, \mathcal{S}, \mathcal{Z})
&= \sum_{i=1}^N\bigg[\log g\left(\by_i(v); \bA_i\bs_i(v), \bm E\right) + \log g\left(\bs_i(v) ; \bs_0(v)+\bbeta(v)'\bm x_i, \bm D\right)  \bigg] \nonumber\\
&+ \log g\left(\bs_0(v); \bm \mu_{\bz(v)}, \bSigma_{\bz(v)}\right) + \sum_{\ell=1}^q \log \pi_{\ell, \bz_\ell(v)},
\end{align}
where $\mathcal{Y} = \{ \boldsymbol{y}_i(v) : i=1,\ldots,N, v = 1,\ldots,V \}$, $\mathcal{X} = \{ \boldsymbol{x}_i : i=1,\ldots,N \}$, $\mathcal{S} = \{ \boldsymbol{s}_i(v) : i=1,\ldots,N, v = 1,\ldots,V \}$, and $\mathcal{Z} = \{ \boldsymbol{z}(v) : v=1,\ldots,V \}$. The model parameters are $\boldsymbol{\Theta} = \{    \{\boldsymbol{\beta}(v)\}, \{\boldsymbol{A}_i \}, \boldsymbol{E}, \boldsymbol{D}, \{ \boldsymbol{\pi_\ell} \}, \{\boldsymbol{\mu_\ell}\}, \{\boldsymbol{\sigma_\ell}^2\} : i=1,\ldots,N, v=1,\ldots,V, \ell=1,\ldots,q\}$. We note that $\boldsymbol{\Theta}$ can be partitioned into  two sets of parameters $\boldsymbol{\Theta}=\{\boldsymbol{\Theta}_{G}, \boldsymbol{\Theta}_{L}\}$.  Here, $\boldsymbol{\Theta}_{G}$ is the set of global parameters which are common across voxels in the brain and include all the parameters in $\boldsymbol{\Theta}$ except $\{\boldsymbol{\beta}(v)\}$, and $\boldsymbol{\Theta}_{L}$ is the set of local parameters which include the voxel-specific covariate effects  $\{\boldsymbol{\beta}(v)\}$.

The EM algorithm requires some initial values for the parameters. HINT obtains these starting values based on estimates derived from an initial analysis using an existing group ICA model via the Group ICA Of fMRI Toolbox (GIFT) \citep{calhoun2001}. GIFT was one of the first group ICA methods developed under the TC-GICA framework for decomposing multi-subject fMRI data. In the iterative steps of the EM algorithm, we obtain the conditional expectation of the complete log-likelihood in the E-step and then obtain the updated parameter estimates in the M-step by maximizing the conditional expectation.  We specify the convergence criteria separately for the global and local parameter estimation given that the two sets of parameter estimates have different convergence properties. The EM-algorithm is presented in Algorithm \ref{algr: EM}.

\begin{algorithm}
\caption{The EM algorithm for estimating the hc-ICA model parameters.}
\label{algr: EM}
	{\bf Initial Values: }{Starting values of $\hat{\boldsymbol{\Theta}}^{(0)}$ and $\hat{\boldsymbol{\beta}}^0$ are obtained using TC-GICA estimates via GIFT.  \\}
	{\bf REPEAT \\}	
	 \hspace{5mm}	{\bf E Step: \\}
		\begin{enumerate}
			\item  Evaluate the conditional distribution $\tilde{p}[ \boldsymbol{s}(v) | \boldsymbol{y}(v); \hat{\boldsymbol{\Theta}}^{(k)} ]$.

			\item Evaluate the conditional expectations in $Q( \boldsymbol{\Theta} |  \hat{\boldsymbol{\Theta}}^{(k)})$ with regard to $\tilde{p}[ \boldsymbol{s}(v) | \boldsymbol{y}(v); \hat{\boldsymbol{\Theta}}^{(k)} ] $ 			
			$$ Q( \boldsymbol{\Theta} |  \hat{\boldsymbol{\Theta}}^{(k)}) = \sum_{v=1}^V E_{\boldsymbol{s}(v) | \boldsymbol{y}(v)}[l_v (\boldsymbol{\Theta}; \mathcal{Y}, \mathcal{X}, \mathcal{S}, \mathcal{Z} )]. $$
			
		\end{enumerate}
	\hspace{5mm}	{\bf M Step: \\}

			\hspace{5mm}  Update parameters as follows
			
			$$\hat{\boldsymbol{\Theta}}^{(k+1)}=argmax_{\Theta} \, Q( \boldsymbol{\Theta} |  \hat{\boldsymbol{\Theta}}^{(k)}) $$
			
{\bf UNTIL } max iterations or  convergence,   i.e.
 $ \frac{ || \hat{\boldsymbol{\Theta}}^{(k+1)}_{G} - \hat{\boldsymbol{\Theta}}^{(k)}_{G} || }{|| \hat{\boldsymbol{\Theta}}^{(k)}_{G} ||} < \epsilon_g$ and $ \frac{ || \hat{\boldsymbol{\Theta}}^{(k+1)}_{L} - \hat{\boldsymbol{\Theta}}^{(k)}_{L} || }{|| \hat{\boldsymbol{\Theta}}^{(k)}_{L} ||} < \epsilon_l.$
			
\end{algorithm}

After obtaining the ML estimates $\hat{\hTh}$, HINT can derive model-based estimates of population- and subject-specific brain networks. In particular, HINT provides model-based estimation of the brain functional networks for specific sub-populations. For a sub-population characterized by covariate pattern $\bx^{\ast}$, the estimated brain functional networks are derived by plugging the ML parameter estimates into the hc-ICA model, i.e.,
\begin{equation}
\label{subpop}
   \hat{\bs}(v)|{\bm x^{\ast}} = \hat{\bs}_0(v) +\hat{\bbeta}(v)^{T}\bm x^{\ast}.
\end{equation}

\subsection{Inference for covariate effects}\label{sec:cov}
In this section, we introduce the statistical inference procedure for testing covariate effects in HINT. Standard maximum likelihood inference is based on the inverse of the information matrix which is used to estimate the asymptotic variance-covariance matrix of the MLEs. However, the information matrix of the hc-ICA model is ultra-high dimensional for brain imaging data and is not sparse, which makes it extremely challenging to invert. HINT implements a novel statistical inference procedure for covariate effects in hc-ICA model developed by our group \citep{shi2016}. The inference method provides an efficient approach to estimate the asymptotic standard errors of the covariate effects at each voxel, i.e., $\hat{\bbeta}(v) (v=1,\ldots,V)$, by directly using the output from the EM algorithm. The method is developed based on the connection between the hc-ICA and linear models. Specifically, the hc-ICA model can be rewritten in a non-hierarchical form by collapsing the multi-levels of models in (\ref{firststage}) and (\ref{secondstage}), which resembles the classical linear model. Therefore, a variance estimator for $\vect\left[\hat{\bbeta}(v)'\right]$ is developed following the linear model theory as,
 \begin{equation}\label{avar}
\mbox{Var}\left\{\vect\left[\hat{\bbeta}(v)'\right]\right\}=\frac{1}{N}\left(\sum_{i=1}^N \bm X_i' \bm W(v)^{-1}\bm X_i\right)^{-1},\end{equation}
where $\bm X_i = \bm x_i'\otimes \bm I _q$ and $\bm W(v)$ is the variance of $\bm \gamma_i(v) + \bm A_i'\bm  e_i(v)$, which is the residual term in the non-hierarchical form of hc-ICA. Then, the variance of $\vect\left[\hat{\bbeta}(v)'\right]$ can be estimated by plugging in an estimator for $\bm W(v)$ in (\ref{avar}), which can be obtained by plugging the ML estimates from the EM algorithm or based on the empirical variance estimator proposed in \cite{shi2016}.

After deriving the variance estimator for covariate effects estimates in hc-ICA, HINT can conduct hypothesis testing on the covariate effects to test for group differences in brain networks. Specifically, users first formulate the hypothesis in terms of linear combinations of the parameters in the hc-ICA model, i.e. $H_0: \bm \lambda' \vect\left[\hat{\bbeta}(v)'\right] = 0 \,\, \mbox{vs.} \,\, H_1: \bm \lambda' \vect\left[\hat{\bbeta}(v)'\right] \neq  0$ where $\bm \lambda$ is a vector of constant coefficients specified based on the hypothesis that users are testing. HINT then constructs the test statistic as.
\begin{equation}
z(v)=\frac{\bm \lambda' \vect\left[\hat{\bbeta}(v)'\right]}{\sqrt{\bm \lambda' \hat{\mbox{Var}}[\vect\left[\hat{\bbeta}(v)'\right]]  \bm \lambda}}.
\end{equation}
The test statistic $z(v)$ is compared against its null distribution to derive the p-value for testing the significance of the covariate effects at voxel $v$.

\section{Toolbox Design}

\begin{figure}[H]
	\caption{A diagrammatic representation of the HINT workflow.}
	\label{fig:flowchartHINT}
	\centering
	\includegraphics[width=0.8\linewidth]{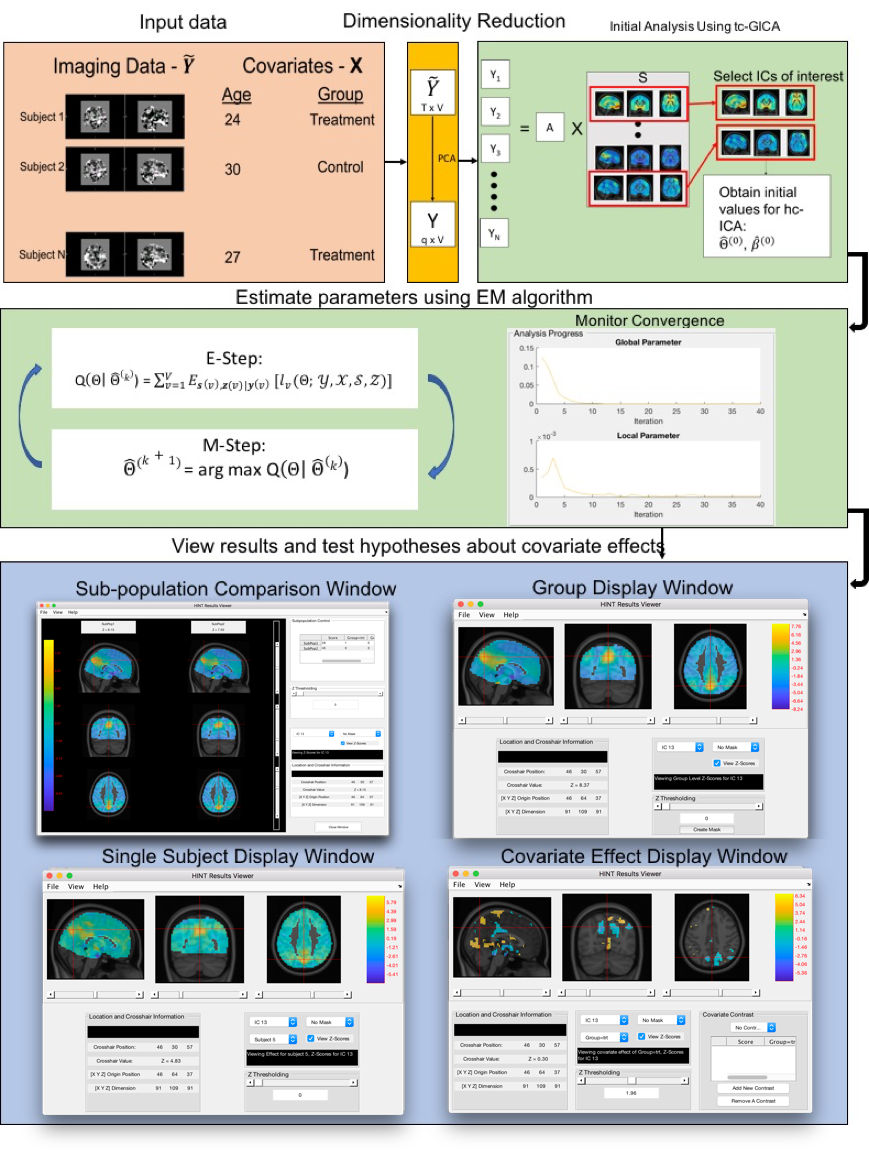}
\end{figure}

The goal of the HINT is to provide a GUI environment that facilitates users to implement the hc-ICA method described in Section 2. Analyses in the HINT can be boiled down to the following steps: loading the imaging data, preprocessing the imaging data, conducting an initial ICA analysis to obtain an initial guess and select ICs of interest (upon user request), estimating the hc-ICA model parameters using the EM algorithm, and visualizing the extracted brain networks on both population- and individual-level and testing hypotheses about covariate effects on brain networks.  Figure \ref{fig:flowchartHINT} provides a schematic flowchart of workflow of the HINT.

Our toolbox design incorporates three major modules implementing the above steps, each with a corresponding panel. The first module allows the user to input and preprocess the data. The second allows the user to run our optimized EM algorithm to estimate the model parameters. The third panel allows visualization of the brain network estimation and testing results. HINT allows users to save their progress in the intermediate analysis steps and load them later if needed to rerun the analysis for some specific steps while avoid repeating the entire analysis. We have included detailed walk through of the HINT toolbox in the Appendix.

To run hc-ICA analysis in HINT, three inputs are required including the preprocessed fMRI images for each subjects, a text file containing the covariates information, and a grey matter mask file. The mask file should be a grey matter mask with either 0s or NaNs in the locations outside the grey matter. These masks can be easily generated using readily available fMRI software such as FSL's FAST tool \citep{zhang2001segmentation}. After providing these inputs, HINT proceeds with the ICA preprocessing procedures as described in Section \ref{subsubsec::preproc}. Specifically, HINT prewhitens the data and generates an initial guess using TC-GICA for the subsequent EM algorithm.

\subsection{Computation of hc-ICA}

Parameter estimation is carried out in the analysis panel via an EM algorithm as described in Section \ref{emAlgs}. The panel provides plots enabling real time monitoring of the algorithm's progress. These plots display the convergence criteria functions for both the global and local parameters in hc-ICA, i.e. $ \frac{ || \hat{\boldsymbol{\Theta}}^{(k+1)}_{G} - \hat{\boldsymbol{\Theta}}^{(k)}_{G} || }{|| \hat{\boldsymbol{\Theta}}^{(k)}_{G} ||} $ and $ \frac{ || \hat{\boldsymbol{\Theta}}^{(k+1)}_{L} - \hat{\boldsymbol{\Theta}}^{(k)}_{L} || }{|| \hat{\boldsymbol{\Theta}}^{(k)}_{L} ||} $, across iterations. The algorithm is terminated if convergence criteria are met or the maximum number of iterations are reached. Please refer to the Appendix for detailed description and screenshot of the analysis panel.

Due to the large dimension of both the fMRI data and the parameter space, parameter estimation for hc-ICA can be quite expensive both in terms of memory and computation. In light of this, extensive testing and care has been put into the EM algorithm implemented in HINT in order to make it as fast as possible, while still allowing it to scale well as the sample size increases. Specifically, we have vectorized the updates for all population-level parameters in the model across data to reduce the computation time. Subject-level parameters are updated individually for each subject which allows the algorithm to scale well without memory issues as the sample size grows. By optimizing the algorithm, HINT has reduced the running time from weeks using the original algorithm in Shi and Guo (2016) to hours. In our experiments, the EM algorithm generally converges within 100 iterations and computation time is in hours with some variations depending on the number of independent components (ICs) and the number of subjects in the analysis.

\subsection{Visualization and Hypothesis Testing}

\begin{figure}
	\centering
	\includegraphics[width=1\linewidth]{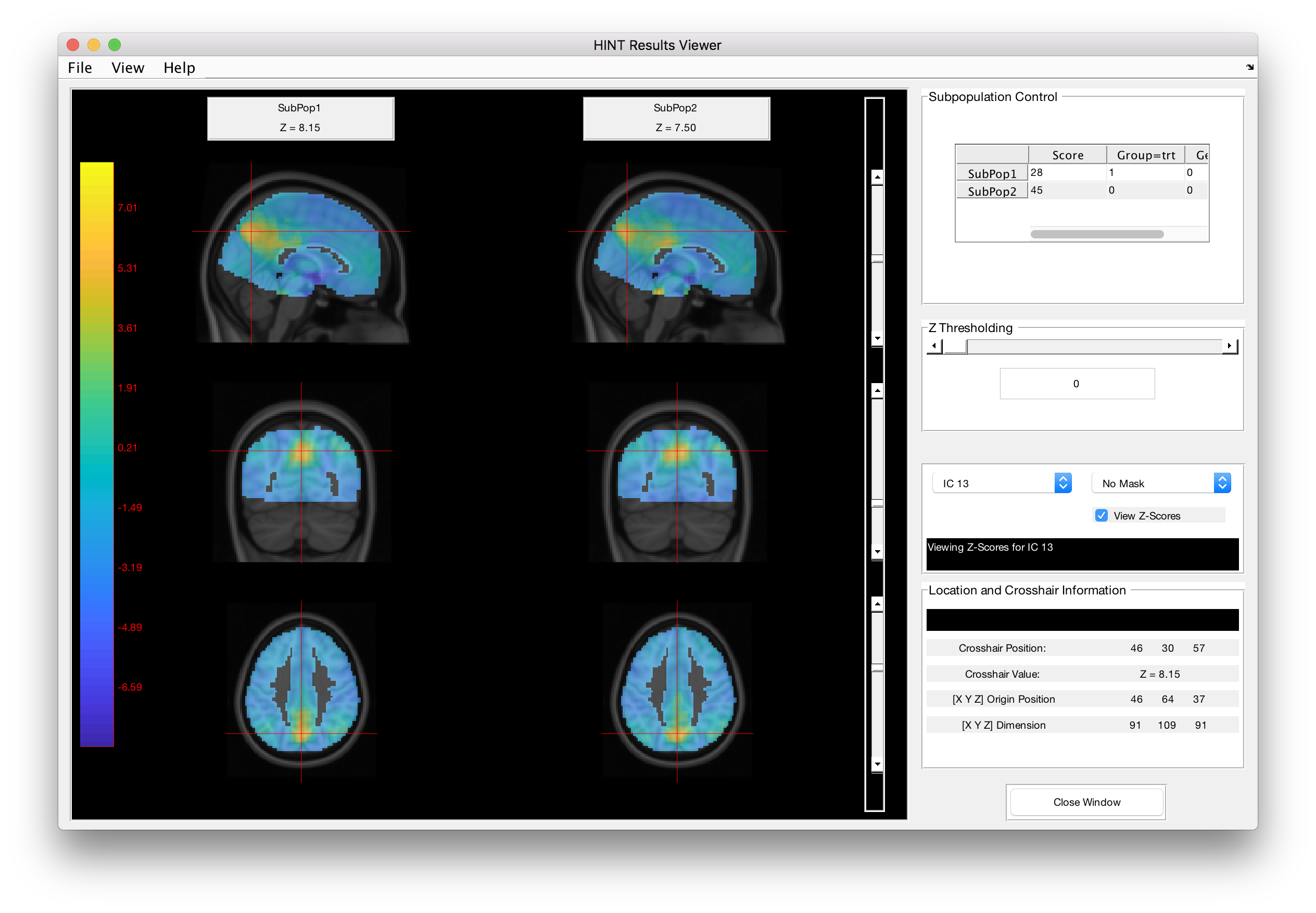}
	\caption{Demonstration of the sub-population comparison window. Sub-populations based on user-defined covariate patterns can be generated and viewed using this display window. The sub-population control box in the upper-right hand corner allows the user to view the corresponding covariate values.}
	\label{fig:subpopCompare}
\end{figure}

The visualization module provides four types of displays, corresponding to the different parameters of interest in the study. The population display shows the estimated population-level brain network maps across all subjects. The sub-population display shows the estimated brain network for specific sub-populations, which are subject groups defined by a specific set of covariate values. The beta display window provides tools for visualization of the estimated covariate effects and hypothesis testing of the effects and contrasts as described in Section \ref{sec:cov}. Finally, the subject-specific display allows examination of individual subject's brain network maps. All panels provide thresholding functionality, and the population display window can be used to generate network masks which can then be applied in the other windows. An example of the sub-population comparison window is provided in Figure \ref{fig:subpopCompare} for illustration. In this example, HINT provides side-by-side comparison of the default mode network maps between two sub-populations defined by user-specified covariates. In the Appendix, we provide a detailed explanation of each visualization window.

\subsection{Command Line Interface}

In addition to the GUI interface, the HINT can also be run using the  command line interface in Matlab. This function facilitates users to reproduce an analysis with a script, conduct simulations studies by running HINT on replicates of synthetic imaging data, or run the HINT analysis on a high performance computing cluster to analyze a large data set. Specifically, the hc-ICA analysis can be conducted using the Matlab function \begin{texttt}{runHINT.m}\end{texttt} from the command line as,

\begin{texttt}{ runHINT(HINTpath, datadir, outdir, q, N, numberOfPCs, maskf, covf,
prefix, maxit, epsilon1, epsilon2) } \end{texttt},

where \begin{texttt}{HINTpath} \end{texttt} is the file path to the HINT toolbox,  \begin{texttt}{datadir} \end{texttt} is the file path to the data directory,  \begin{texttt}{outdir} \end{texttt} is the file path to the output directory,  \begin{texttt}{q} \end{texttt} is the number of independent components,  \begin{texttt}{N} \end{texttt} is the number of subjects,  \begin{texttt}{numberOfPCs} \end{texttt} is $R$ which is the number of principal components extracted in the two stage dimension reduction of the initial TC-GICA analysis via GIFT,  \begin{texttt}{maskf} \end{texttt} is the file path for the mask file in Nifti format,  \begin{texttt}{covf} \end{texttt} is the file path for the .csv file containing the covariates,  \begin{texttt}{prefix} \end{texttt} is the desired prefix for the output,  \begin{texttt}{maxit} \end{texttt} is the maximum number of EM iterations, and  \begin{texttt}{epsilon1} \end{texttt} and  \begin{texttt}{epsilon2} \end{texttt} are the convergence criteria $\epsilon_g$ and $\epsilon_l$ described in detail in the Appendix. After the script finishes running, the results from the hc-ICA can be viewed through the HINT visualization window.

\section{Comparison to Some TC-GICA Approaches}

\begin{figure}
	\centering
	\includegraphics[width=1.0\linewidth]{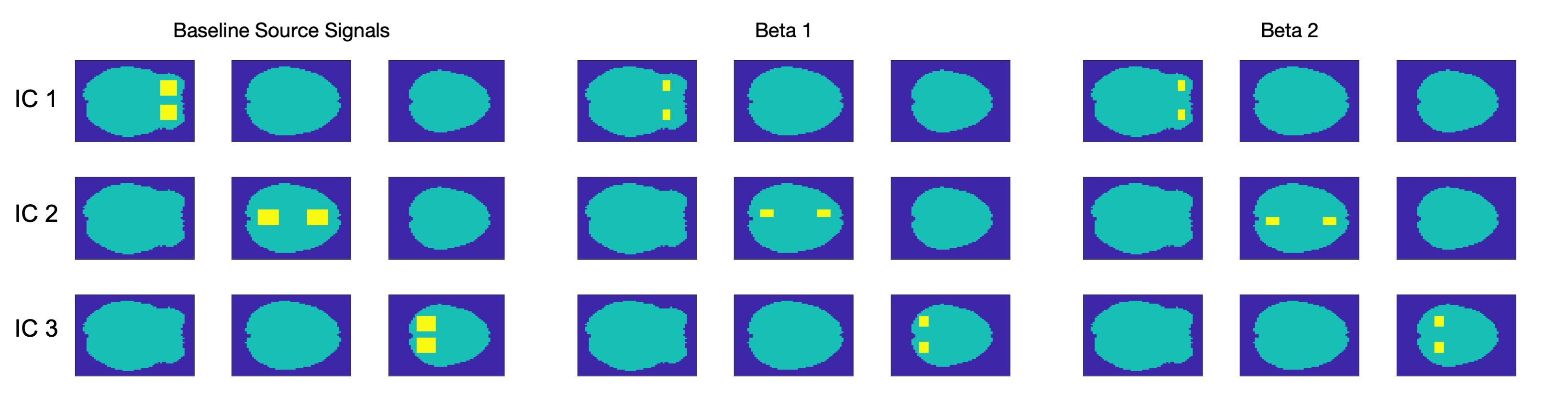}
	\caption{Spatial maps used for the simulation (non-overlapping $S_0$ and non-overlapping covariate effects).}
	\label{fig:simulation_data_ex_nooverlap}
\end{figure}

We conduct a comparative study between the HINT and three existing TC-GICA methods using synthetic data to illustrate its performance in detecting network differences and estimating spatial maps. 
Following \cite{shi2016}, we generated synthetic fMRI datasets from three underlying source signals ($q=3)$ and considered various sample sizes with $N = 25, 50, 100$ and $200$. For each source or IC, we first generated 3D population-level spatial maps, i.e. ${s_0(v)}$, with dimension $53\times63\times3$ based on three selected slices from real fMRI imaging data. We then generated two covariates for each subject with one ($x_{i1}$) being a categorical covariate simulated from Bernoulli$(0.5)$ distribution and the other ($x_{i2}$) being a continuous covariate randomly generated from a uniform $(0, 1)$ distribution. The strengths, $\bm {\beta}(v)$, for these voxels were taken  from $\{2.0, 3.0, 4.0\}$. In this paper, we generated synthetic data from more complex settings where the spatial sources and covariate effects can be overlapping. Specifically, we considered all four possible scenarios with the combinations of overlapping or non-overlapping source signals and overlapping or non-overlapping covariate effects. Figure \ref{fig:simulation_data_ex_nooverlap} shows the activated regions in each IC and the covariate effect maps for the non-overlapping case. Highlighted regions correspond to active voxels. A full view of all combinations of overlapping and non-overlapping maps can be found in Table \ref{fig:allmaps} of Appendix 3.

The subject-specific spatial source signals were simulated as the linear combination of the population-level signals, covariate effects and subject-specific random effects. We considered three levels of between-subject variability, labeled as ``low",  ``medium," and ``high," which correspond to between-subject variances of $\text{diag}\{0.1, 0.3, 0.5\}$, $\text{diag}\{1.0, 1.2, 1.4\}$, and $\text{diag}\{1.8, 2.0, 2.5\}$, respectively. For temporal responses, each IC had a time series of length of T = 200 that was generated based on time courses extracted from real fMRI data and hence represented realistic fMRI temporal dynamics. We then generated subject-specific time sources with similar frequency features but different phase patterns \citep{guo2011}, which represented temporal dynamics in resting-state fMRI signals. After obtaining the spatial maps and time courses for the source signals, Gaussian background noise was added to generate observed fMRI data. Table \ref{simfactors} displays a summary of the simulation settings we varied. We considered all combinations of settings and generated 50 datasets per combination.

\begin{table}
\caption{Factors varied in the simulation study. For each combination of factors we generated data from 50 datasets.}
\label{simfactors}
\centering
\begin{tabular}{l|c} \hline
Factor & Levels \\ \hline
Overlapping $s_0$ maps & no, yes\\
Overlapping $\beta$ maps & no, yes \\
$N$ & 25, 50, 100, 200 \\
between-subject variability & low, medium, high\\
\hline
\end{tabular}
\end{table}

%
%
\medskip
\noindent{\underline{Comparison Methods}} \\

We compared the performance of HINT with three Group ICA methods: group ICA with GICA3 back reconstruction (GICA3) \citep{erhardt2011comparison} and group ICA with the spatio-temporal back reconstruction (STR) , as implemented in the GIFT toolbox \citep{calhoun2001}. We also compared to group information guided ICA (GIG-ICA) using the GIFT-estimated group-level spatial maps as the reference signals \citep{du2013group}, which was  was implemented using the gig-ica package available on NITRIC\footnote{\url{https://www.nitrc.org/projects/gig-ica/}}. For the GICA3, GIG-ICA, and STR approaches, once we have obtained back-reconstructed subject-level spatial maps, we fit the following regression model at each voxel and IC (denoted $q$),

\begin{equation}
    s_{iq}(v) = s_{0q}(v) + \beta_{1q}(v) x_{i1} + \beta_{2q}(v) x_{i2} + \epsilon_{iq}(v),
\end{equation}

\noindent for $q = 1,2,3$. Significance was assessed using the p-values for the estimated coefficients.



\medskip
\noindent{\underline{Performance Metrics}} \\

We estimated the false positive rate or Type-I error rate with the empirical probabilities of falsely detecting covariate effects at voxels such that there were no covariate effects, i.e. $\beta(v)=0$. We estimated the power of the tests with the empirical probabilities of detecting covariate effect at voxels with non-zero covariate effects, i.e., $\beta(v) \neq 0$. We report the correlation between the true and estimated population-level spatial maps and the true and estimated subject-specific maps in order to measure how well each approach does in estimating the population- and individual-level spatial maps.

\medskip
\noindent{\underline{Results}} \\

\begin{figure}[H]
	\centering
	\includegraphics[width=0.95\linewidth]{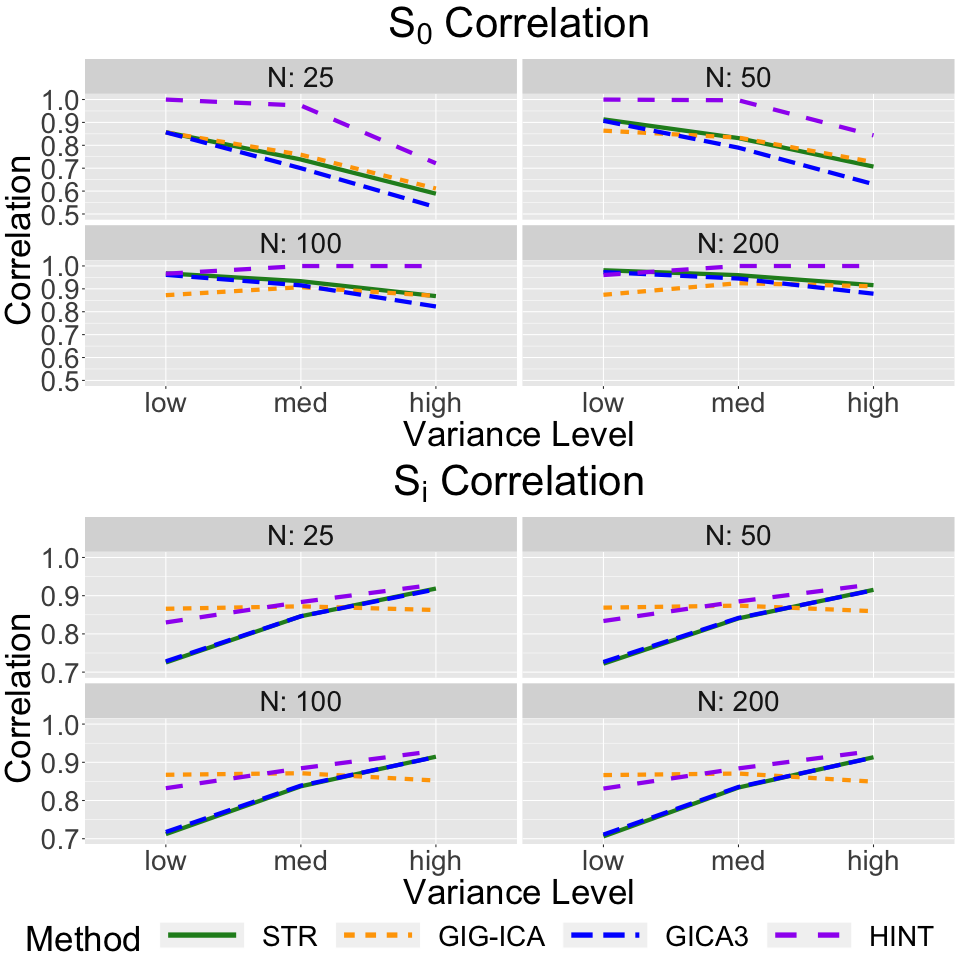}
	\caption{Correlation with the true population-level and individual-level spatial maps with STR, GICA3, GIG-ICA, and HINT for the synthetic data with no spatial overlap in the baseline maps or the covariate effect maps.}
	\label{fig:so0bo0_cor}
\end{figure}

\begin{figure}[H]
	\centering
	\includegraphics[width=0.95\linewidth]{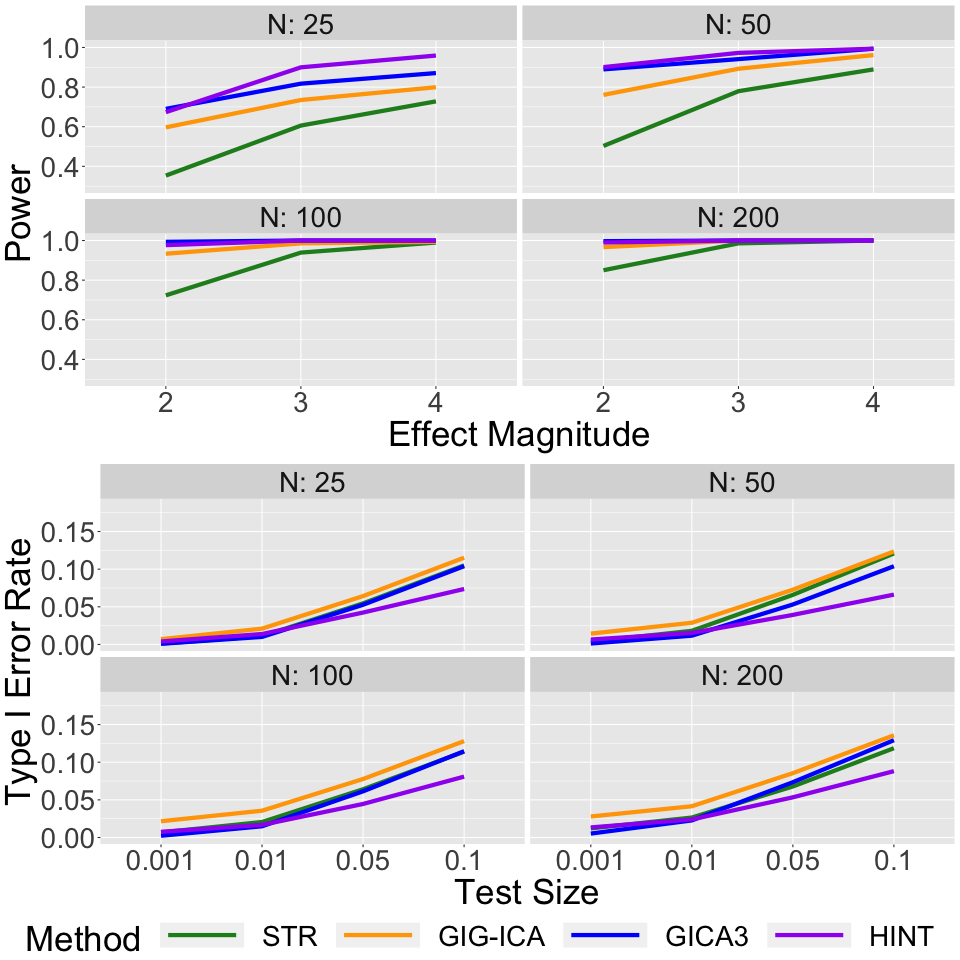}
	\caption{Power and Type I Error Rates for testing covariates' effects with STR, GICA3, GIG-ICA, and HINT for the synthetic data with no spatial overlap in baseline source signals or covariate effect maps.}
	\label{fig:so0bo0_powert1}
\end{figure}

In general, we found the performance of HINT is very similar across the the four scenarios with various combinations of overlapping or non-overlapping signals. It only shows slight decrease of accuracy in estimating the population-level spatial maps in the presence of overlapping signals. We show the results for the scenario with non-overlapping source signals and non-overlapping covariate effect maps in Figures \ref{fig:so0bo0_cor} and \ref{fig:so0bo0_powert1}. The results for the other three configurations with overlapping in either source signals or covariate effects are presented in Figure 15--20 in the Appendix, which are generally consistent with the results from the non-overlapping setting. The simulation results in Figure \ref{fig:so0bo0_cor} show that HINT generally demonstrates better or similar accuracy in recovering both the population-level IC spatial maps (i.e. $S_0$) and the subject-level IC spatial maps (i.e. $S_i$), as compared to the other approaches. For testing covariates' effects on brain networks, the results in Figure \ref{fig:so0bo0_powert1} show that with small sample size $N=25$ the HINT approach has the highest power to detect true covariate effects, followed by  GICA3, GIG-ICA, and then STR. As the sample size increases, GICA3, GIG-ICA, and HINT have similar statistical power, all of which outperform the STR approach. The Type I error rates are generally comparable between STR, GIG-ICA and GICA3, while HINT shows a lower Type I error rate.
 
\section{Conclusions}\label{sec:summary}

In this paper we introduced the HINT Matlab toolbox for implementing hierarchical-covariate adjusted ICA (hc-ICA) \citep{shi2016} which is the first group ICA method that models covariate effects in the ICA decomposition. HINT enables users to perform hierarchical ICA analyses using an intuitive and user friendly interface, that neatly separates data loading/ preparation, algorithm execution and result visualization using separate tabs and windows. HINT provides interactive visualizations of the output of analyses aiding users in testing hypotheses about covariate effects on brain networks. HINT also offers a script based execution work flow for non-gui server environment users. HINT is useful for both statisticians and neuroscientists for investigating differences in brain functional networks between clinical sub-populations while controlling for potential confounding factors. To the best of our knowledge, HINT will be the first ICA toolbox that allows users to conduct hypothesis testing on user-specified linear contrasts of covariate effects and to obtain model-based prediction of sub-population brain network maps based on user-specified characteristics. While this paper focused on the implementation of the hc-ICA model using the HINT toolbox, related ICA methods developed under the hierarchical ICA  framework are going to be added to the HINT. For example, we have recently developed a longitudinal ICA model \citep{wang2019hierarchical} for modeling longitudinal fMRI data and this method will be incorporated into the future version of the HINT toolbox.

The first level model of the hc-ICA model in equation (2) of the paper bears a resemblance to the general linear model (GLM). However, there are several fundamental differences between hc-ICA and GLM: 1) unlike the GLM which is based on a pre-specified design matrix, hc-ICA doesn't require any prior information in it model where both the mixing matrix $\bm{A}$ and spatial source signals $\bm{s}$ are unknown and need to be estimated based on the observed imaging data, 2) the hc-ICA is a multi-level model which includes the first-level and second-level models in (2) and (3) and also the Mixture of Gaussian (MoG) source distribution model in (4). In comparison, the GLM is a single-level model that doesn't model the various levels of variability and doesn't model the unobserved latent source signals $\bm{s}$, 3) the estimation methods are fundamentally different between hc-ICA and the GLM. hc-ICA uses an EM algorithm to simultaneously estimate all the parameters and latent variables across its multi-level models while the GLM mainly uses least square estimation given its relatively simple model set up.

When applying ICA to extract brain networks, we often need to determine whether the identified independent components (ICs) are related to neural processing or caused by non-neural related artifacts. In its current form, HINT includes a feature that allows supervised removal of artifactual ICs after obtaining an initial guess by visual inspection of the spatial maps of the extracted ICs or by correlating the IC time series with physiological related temporal signals. In the future, we plan to add an automated procedure, similar to the FIX algorithm \citep{salimi2014}, into our HINT toolbox to help identify and remove nuisance ICs. Specifically, a classifier could potentially be developed to calculate the predicted probability that an IC is artifactual. Such a classifier could help guide the decision of whether to include or remove an IC from the data.

HINT uses estimates from TC-GICA to provide the starting values to initiate the EM algorithm.  A limitation of the EM algorithm is the potential for converging to a local maximum for the objective function. For a fixed initial guess, the EM algorithm is theoretically guaranteed to converge to a maximum, but not necessarily the global maximum. One commonly used strategy to overcome this limitation of EM is to consider different starting values, e.g. by adding random variations to the initial TC-GICA estimates and re-run the estimation procedure multiple times. We have adopted this strategy in our previous work \citep{shi2016, wang2019hierarchical}.

\section{Acknowledgements}

Research reported in this publication was supported by the National Institute Of Mental Health of the National Institutes of Health under Award Number R01MH105561 and R01MH118771 and by the National Center for Advancing Translational Sciences of the National Institutes of Health under Award number UL1TR002378. The content is solely the responsibility of the authors and does not necessarily represent the official views of the National Institutes of Health.

\bibliographystyle{apalike}
\bibliography{hcICAreferences}

\newpage

\section*{{ Appendix 1 - Toolbox Details}}

In the following sections, we walk through the main steps of the HINT toolbox.  To demonstrate the functionality of the toolbox, we perform an entire hc-ICA analysis using a fMRI dataset with 24 subjects. We have measures of three covariates: score on a behavioral test, treatment group, and gender. Our primary interest is in identifying a treatment effect on the spatial maps, however we also are interested in any possible differences in brain networks due to gender or score. The steps of the analysis will be carried out using the following 3 panels:

\begin{enumerate}
	\item The prepare analysis panel (``Prepare Analysis") is where the user inputs the data, performs model specification of the covariates and interactions, conducts initial TC-GICA analysis to obtain initial values for the hc-ICA EM algorithm, and selects ICs of interest for hc-ICA modeling if requested by the user.
	\item The analysis panel (``Run analysis") is where the user applies the EM algorithm to estimate the hc-ICA model parameters. The user can specify the maximum number of iterations and the convergence criteria for the EM algorithm, as well as monitor the convergence of the EM algorithm estimates using two change plots displaying the changes in the global and local parameters across EM iterations.
	\item The results display panel (``Visualize") provides visualization GUIs for the results from the hc-ICA model analysis including the model-based estimates of the population- and individual-level spatial maps for brain networks and the estimated covariate effects maps. The visualization GUI also allows users to specify covariate patterns for a sub-population of interest and obtain the model-based estimation/prediction of the corresponding brain network. The covariate viewer within the visualization GUI enables model-based hypothesis testing about covariate effects following the procedure outlined in Section 2.4.
	
\end{enumerate}

\subsection{Prepare Analysis Panel}
\begin{figure}[H]
	\centering
	\includegraphics[width=1\linewidth]{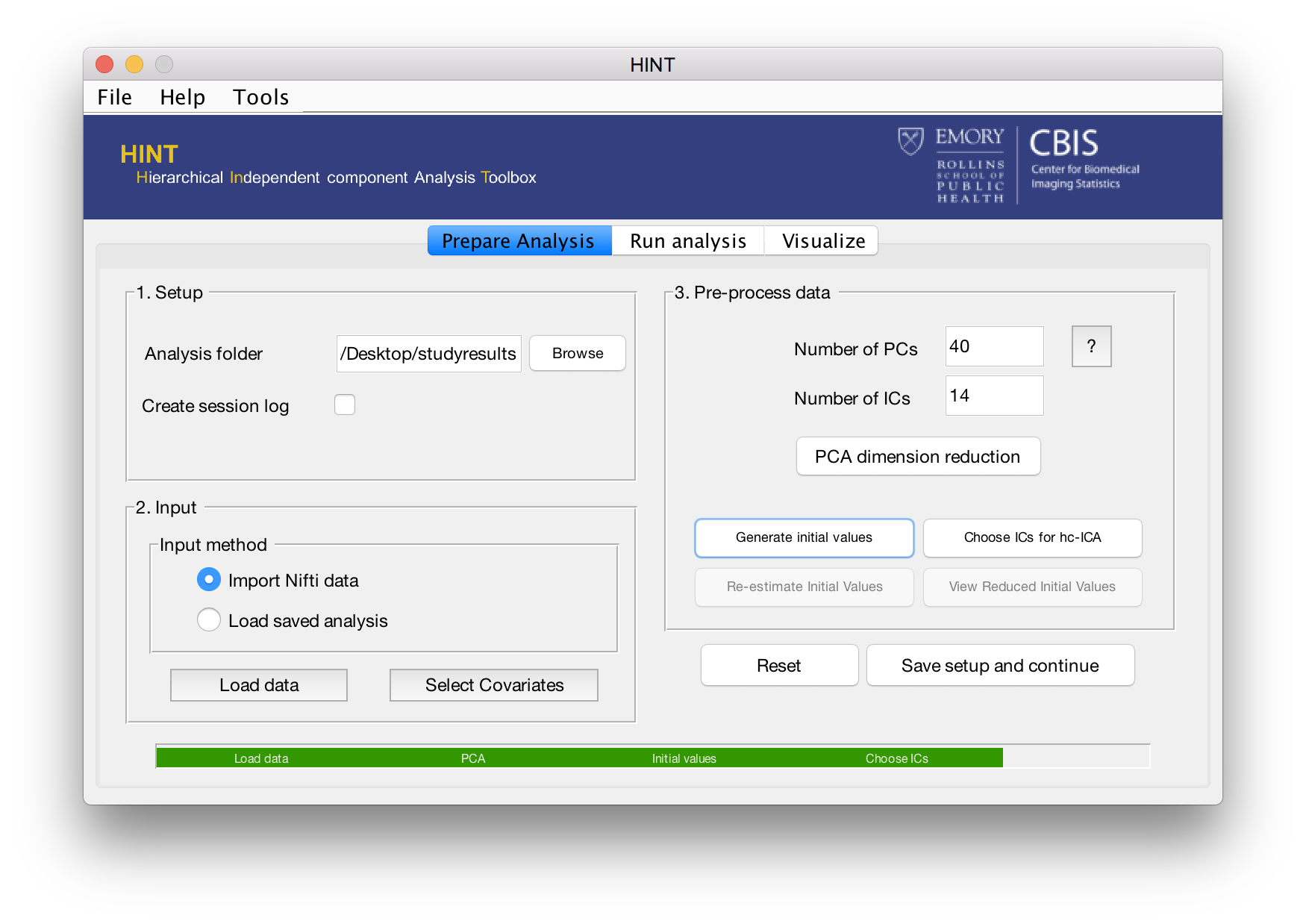}
	\caption{The HINT analysis preparation panel in which the user can setup their analysis and obtain initial values for the EM algorithm for the hc-ICA model parameters.}
	\label{fig:preprocessingpanel}
\end{figure}

In the Prepare Analysis panel (``Prepare Analysis", Figure \ref{fig:preprocessingpanel}), users can input the data, perform model specification, preprocess the images using the method in Section 2.1,  conduct initial analysis using a TC-GICA model to generate initial values for the hc-ICA EM algorithm, and select ICs of interest for the subsequent hc-ICA modeling.

\subsubsection{Data Input}\label{ssec:datainput}

The first section, i.e. ``1. Set up", of the panel asks the user to specify the folder for the analysis, along with selecting whether they would like to generate an output log. All output from the HINT is stored in the selected folder. If the output log is selected, a text file will be created in the analysis folder that provides information about the preprocessing and analyses performed.

In ``2. Input",  the user has two options when inputting data for the analysis. They can start a new analysis by selecting ``Import Nifti files" and inputting the images in Nifti format. Alternatively, if the user wishes to continue an analysis that they have already started, they can load the runinfo file (described later)   corresponding to that analysis using the ``Load saved analysis" option, allowing them to bypass the preprocessing and initial analysis.

When starting a new analysis, the user is instructed to provide three elements: the Nifti files containing the subject-level data, a mask file in Nifti format, and a file containing the subjects' covariates. The mask file should be a grey matter mask with either 0s or NaNs in the locations outside the grey matter. These masks can be easily generated using readily available fMRI software such as FSL's FAST tool \citep{zhang2001segmentation}. The covariate file must be a .csv file conforming to the following structure. The top row of the covariate file contains variable names where the first variable name needs to be specified as ``subject" since the first column is reserved for subjects' filenames.  Then, each of the following rows contains the covariates of a subject. In particular, the first column includes the filename for the subject file. An example covariate file layout is provided in Table \ref{ex:covfile}. In this file, we included the three covariates: score, treatment group, and gender.

\begin{table}[]
	\centering
	\begin{tabular}{lccc}
		\hline
		subject      & Score & Group      & Gender \\ \hline
		subj1.nii  &  28  & Trt   & 1 \\
		subj2.nii  & 36 & Trt  & 1 \\
		$\vdots$  &$\vdots$    & $\vdots$ & $\vdots$ \\
		subj5.nii  & 42 & Ctrl    & 0 \\
		\hline
	\end{tabular}
    \caption{An example covariate file.}
   	\label{ex:covfile}
\end{table}

\begin{figure}[H]
	\centering
	\includegraphics[width=1\linewidth]{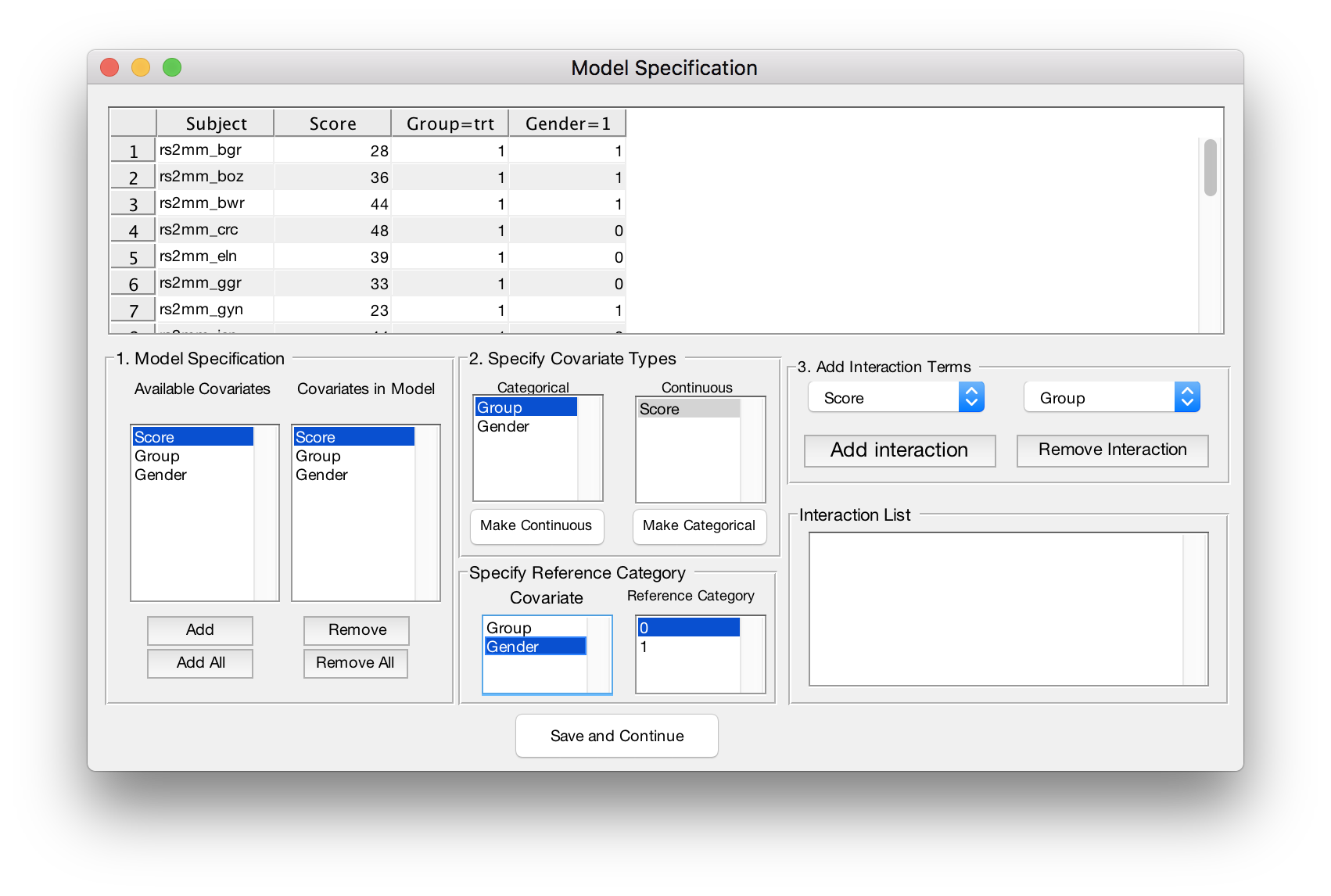}
	\caption{The covariate display window. Here the user can view the design matrix, verify that covariates are coded correctly, change covariate traits (``continuous" vs. ``categorical"), change the reference group for categorical covariates, and add interaction effects.}
	\label{fig:covwindowex}
\end{figure}

After reading in the data, the user has the option to perform model specification using the ``Select Covariates" button (Figure \ref{fig:covwindowex}). Pressing this button opens up the ``Model Specification" window in Figure \ref{fig:covwindowex}. This window provides users with the option to include and exclude individual covariates from the analysis, specify interactions to include in the analysis, and to specify whether covariates are continuous or categorical. By default, the HINT treats string values (e.g. ``Group" in Table \ref{ex:covfile}) as categorical covariates and performs reference cell coding, creating binary indicators for all the non-reference levels. Numeric values are treated as continuous covariates by default (e.g. ``Score" in Table \ref{ex:covfile}). If the user has integer codings for categorical covariates (e.g. ``Gender" in Table \ref{ex:covfile}) and would like to treat the integer values as categorical levels, they can switch the covariate from ``continuous covariates" to ``categorical covariates" using the ``Specify Covariate Types" panel shown in Figure \ref{fig:covwindowex}. Finally, the reference category for categorical covariates can be changed using the ``Specify Reference Category" boxes. For example, in Figure \ref{fig:covwindowex} we have changed the reference group to be the control group. 

\subsubsection{Initial TC-GICA analysis}\label{subsubsec::iniguess}

The EM algorithm used to estimate the parameters in the hc-ICA model functions most efficiently when provided with a good set of starting values. To obtain the set of starting values, HINT first conducts an initial analysis using GIFT \citep{calhoun2001}. In addition to providing a reasonable set of starting values, the initial analysis can also help users identify ICs of interest for the hc-ICA model. For example, when performing ICA of fMRI data, there may be some ICs that are not neural-related but correspond to artifacts including motion effects. It is possible to remove these artifact ICA components from the data and then perform the ICA \citep{salimi2014, griffanti2014}. Furthermore, among the extracted ICs, users may have strong interests in specific brain functional networks and would like to focus on these networks in hc-ICA modeling. After running the initial analysis, HINT dispalys the estimated ICs from GIFT and provides users the option of selecting a subset of ICs of interest for the subsequent hc-ICA modeling while removing the rest of the ICs from the data.

The initial analysis via GIFT consists of several steps. First, a two stage dimensionality reduction step is performed prior to TC-GICA.  At the first stage, each subject's data is reduced to $R$ principal components. Then, the PCA-reduced subject data is stacked to create an $NR \times V$ data matrix where $N$ is the number of subjects and $V$ is the number of voxels in the brain mask. The second-stage dimension reduction is then performed on this stacked group matrix to reduce it to a $q \times V$ data matrix where $q$ is the number of independent components to be extracted.  To implement this two-stage dimension reduction, we specify $R$ and $q$ in the ``Number of PCs" and ``Number of ICs" boxes in the preprocessing panel (Figure \ref{fig:preprocessingpanel}). After the second stage of reduction, spatial ICA is performed on the reduced data to obtain the $q \times V$ independent components corresponding to population-level brain networks and the corresponding mixing matrix. Next, subject-specific ICs are obtained using the back reconstruction approach as described in \cite{calhoun2001}. Based on these subject-specific IC estimates, HINT generates the initial values for the parameters in the hc-ICA model by clicking the ``Generate initial values" button (Figure \ref{fig:preprocessingpanel}). Users can then visualize the estimated IC maps from the TC-GICA initial analysis and, if they want, select ICs of interest for the subsequent hc-ICA modeling using the ``Choose ICs for hc-ICA" button (Figure \ref{fig:preprocessingpanel}).

\subsubsection{Saving the analysis setup via the runinfo file}\label{subsubsec::runinfo}
In applications, it is often desirable to save the analysis setup to facilitate reproducing the analysis on a later occasion or to re-perform the analysis with some modifications. HINT enables this reproducibility by building an information file entitled the ``runinfo" file containing the information about the data set and the analysis set up based on the specifications from the current GUI. Table \ref{runinfoTable} in the Appendix displays the list of objects included in ``runinfo" file. This file is automatically created when the user selects the ``Save setup and continue" option displayed in Figure \ref{fig:preprocessingpanel}. When clicked, this button asks the user to specify a prefix for the analysis. A subfolder is created with this prefix, and the runinfo file is written to this folder, along with all EM algorithm output. This file can be referenced when the user chooses to repeat the analysis on the same data on a later occasion and eliminates the need to repeat some of the steps such as the loading the Nifti files and preprocessing of the images. The runinfo file is a single file, and thus can be easily moved across different computers or networks.

\subsection{The Analysis Panel}

After performing model specification and obtaining an initial guess for the model parameters, the second panel (Figure \ref{fig:analysispanel}) is used to carry out the hc-ICA analysis, estimating parameters in the specified model via EM algorithm. HINT provides two EM iteration plots to monitor the convergence of the EM estimates for the global parameters and local parameters, respectively. The analysis panel in HINT allows users to specify the following parameters for EM convergence:

\begin{figure}[H]
	\centering
	\includegraphics[width=1\linewidth]{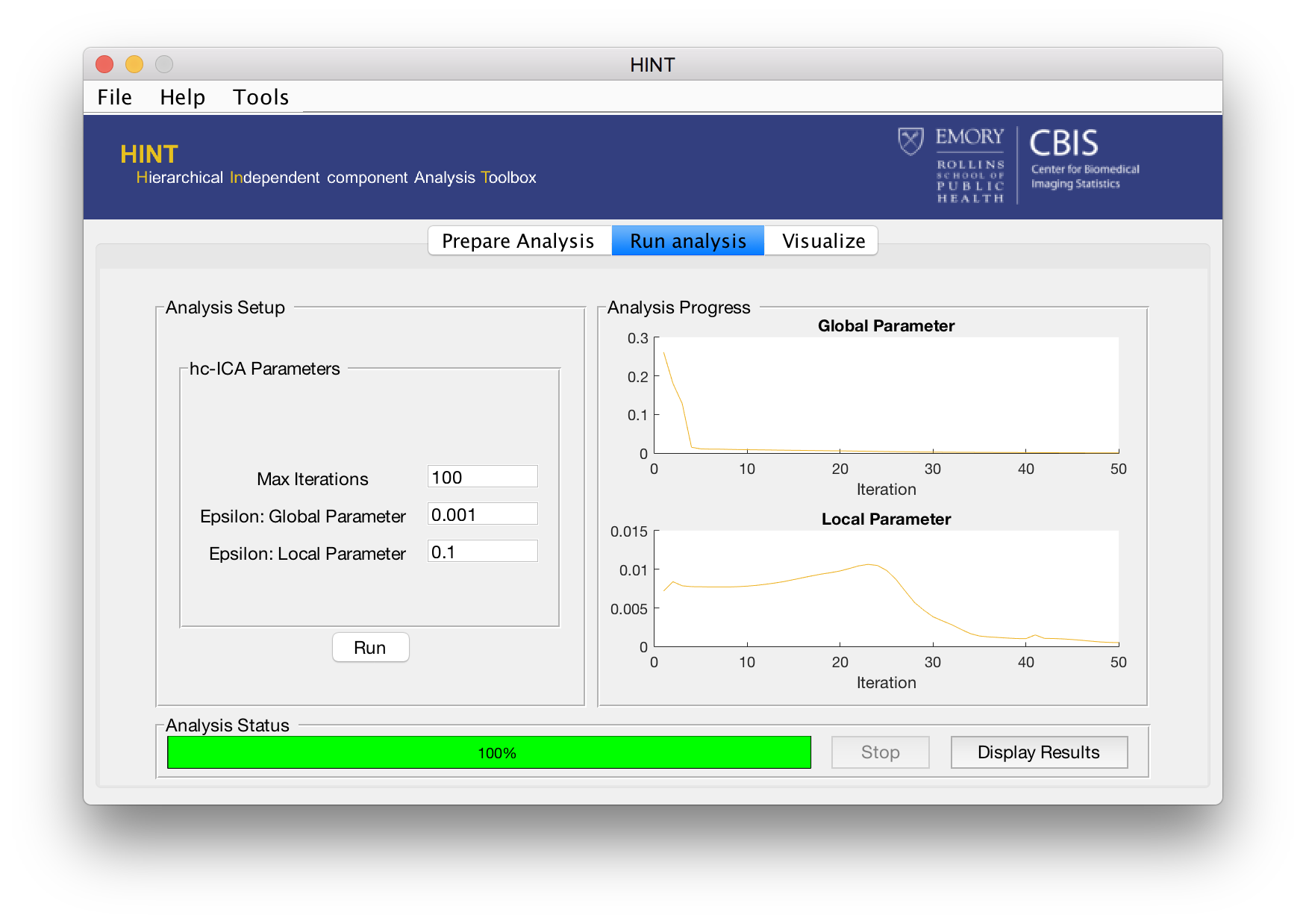}
	\caption{The HINT analysis panel. This panel allows the user to select settings for the estimation procedure and to track the progress of the algorithm via EM iteration plots.}
	\label{fig:analysispanel}
\end{figure}

\begin{description}
	\item [Max Iterations] The maximum number of EM algorithm iterations.
	\item [Epsilon: Global Parameters] The termination criterion for the convergence of the global parameters, i.e. $\epsilon_g$ in  Algorithm 1.
	\item [Epsilon: Local Parameters]  The termination criterion for the convergence of the local parameters, i.e. $\epsilon_l$ in  Algorithm 1.
	\end{description}

The EM algorithm stops when the convergence criteria are met or when reaching the maximum number of iterations. After selecting ``Run", the user can monitor the algorithm's progress using the iteration plots displayed on the right hand side of the analysis panel (Figure \ref{fig:analysispanel}). These two figures display  $ \frac{ || \hat{\boldsymbol{\Theta}}^{(k+1)}_{G} - \hat{\boldsymbol{\Theta}}^{(k)}_{G} || }{|| \hat{\boldsymbol{\Theta}}^{(k)}_{G} ||} $ and $ \frac{ || \hat{\boldsymbol{\Theta}}^{(k+1)}_{L} - \hat{\boldsymbol{\Theta}}^{(k)}_{L} || }{|| \hat{\boldsymbol{\Theta}}^{(k)}_{L} ||} $ across iterations. Additionally, through the iteration plots, HINT offers the user the flexibility of manually stopping the algorithm when it is needed. This flexibility is useful in practice because the EM convergence criteria may need to vary across different datasets due to the the differences in the dimension of parameters and sample sizes and the user may lack sufficient information in pre-specifying a suitable convergence criteria for a particular dataset. By monitoring the iteration plots, if at some point the user is already satisfied with the convergence, they can manually terminate the EM algorithm by using the ``Stop" button even though the pre-specified termination criteria have not yet been met. In this case, the algorithm will terminate at the end of the current iteration and output the results. Convergence time will vary based on the dataset, the number of independent components, and the number of subjects in the analysis. However, in general we find that the approach converges within several hours.

After the algorithm terminates, either by reaching the stopping criteria or by user intervention, HINT saves all relevant results. These files can be found in the output subfolder of the directory specified in the ``prepare analysis" window. All files will begin with the user-specified prefix. The following items are saved: the overall population aggregate maps, the $\bm{s_0}$ maps, the covariate effect maps, and maps of the estimated standard errors for the covariate effects. Individual subject level results can be found in the iteration results file for the final completed iteration.

\subsection{Visualization Panel} \label{visPanel}

\begin{figure}[H]
	\centering
	\includegraphics[width=1\linewidth]{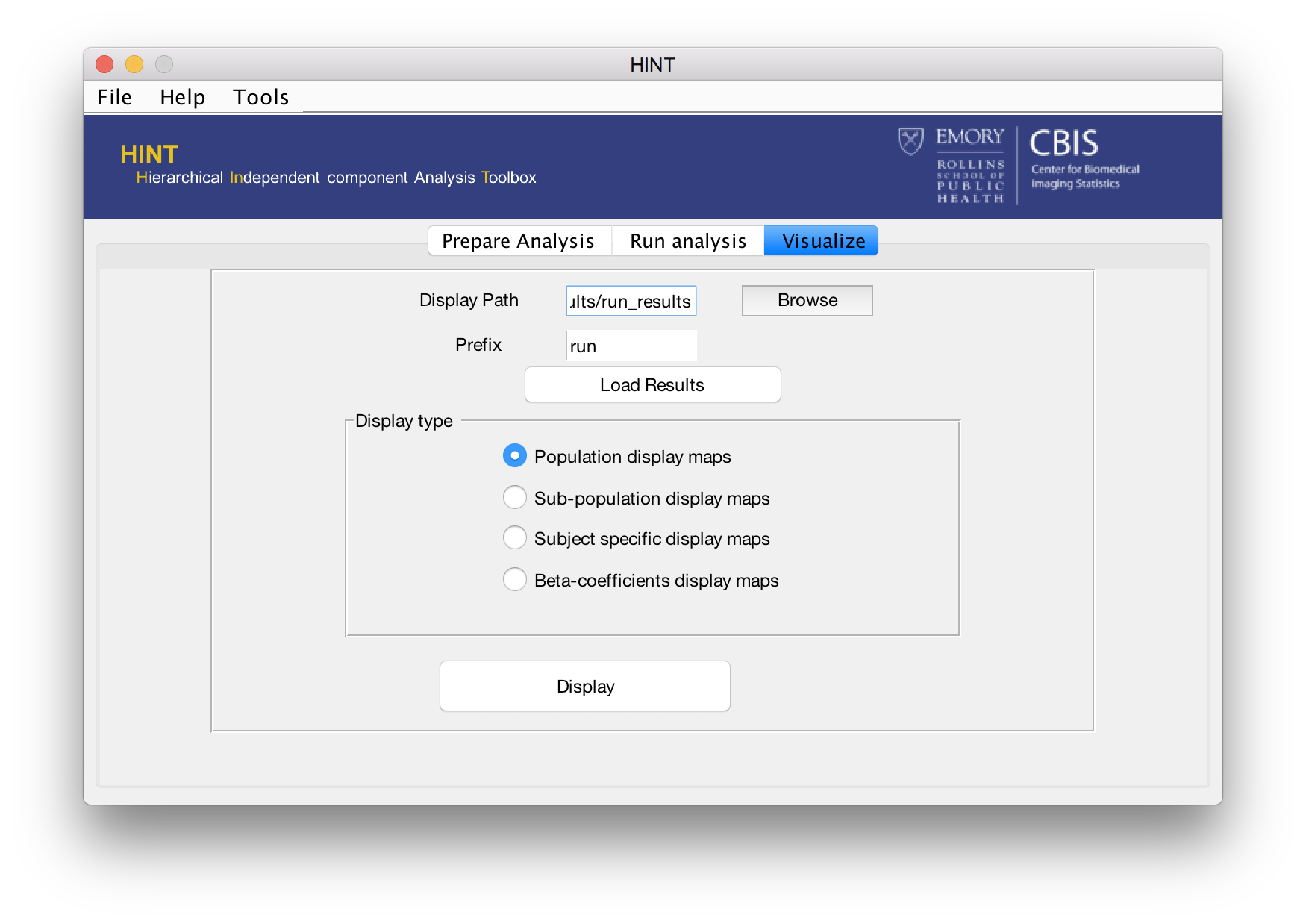}
	\caption{The HINT visualization panel. In this panel the user can view the overall study population-averaged maps, compare sub-populations, look at the estimated ICs for individual subjects, and examine the estimated beta coefficient maps.}
	\label{fig:visualizationpanel}
\end{figure}

The third and final panel in HINT is the visualization panel. Figure \ref{fig:visualizationpanel} displays the panel, which enables the user to visualize the results from the hc-ICA analysis overlaid on the brain. There are two ways that the user can visualize results. First, if the user has just completed an hc-ICA analysis and is coming from the second HINT panel, then the results will already be loaded and they can proceed directly to selecting a viewer window. Second, if the user wishes to view results from a previously completed analysis, they can fill out the display path and prefix boxes using the options they selected in the prepare analysis panel when they first performed the analysis.

The user can select the type of result to view in the ``Display type" section of the visualization panel. HINT provides display viewers for the following main results from hc-ICA: ``Population display maps" shows the model-based estimates of the population-level brain network maps,  ``Sub-population display maps" shows model-based estimates of the brain network maps for particular sub-populations defined by user-specified covariate patterns,  ``Subject specific display maps" shows model-based estimates of network maps for individual subjects in the data, ``Beta-coefficients display maps" shows the estimated covariate effects including the maps of the beta parameters, i.e. $\{\hat{\boldsymbol{\beta}(v)}\}$, and also maps of user-specified linear combinations of the beta parameters.   In each of the four display viewers, the user is able to move around the maps with the mouse cursor to view different locations of the brain network.  There is a ``Location and Crosshair Information" section in each of the display viewers, adapted from the BSmac viewer \citep{zhang2012}, that provides the user with information on the current brain location at the cursor. Specifically, the user can view the coordinates of the current crosshair in the 3D brain image, the corresponding Brodmann area the current crosshair is located in, and the value of the displayed estimates or its corresponding Z-score at that crosshair. Next, we demonstrate the four display viewers using the fMRI study data.

\subsubsection{Population average display maps}

\begin{figure}[H]
	\centering
	\includegraphics[width=1\linewidth]{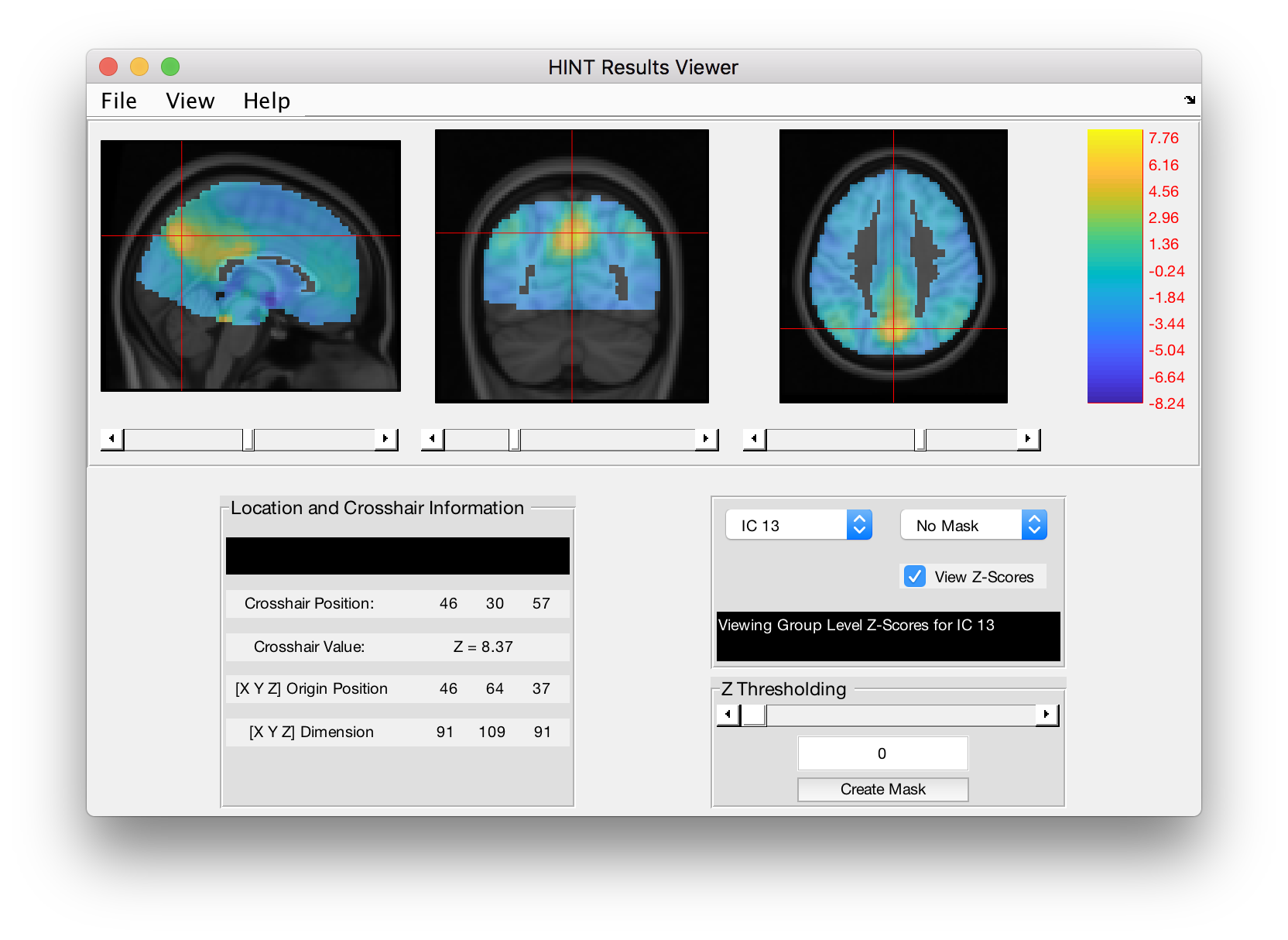}
	\caption{The HINT study population-level display window. Here the user can view the population average maps, as well as create masks using the ``create mask" button in the lower right-hand panel. These masks can be applied in other windows such as the single subject viewer.}
	\label{fig:groupViewer}
\end{figure}

 The first display window, the study population level display, allows the user to view the population-level brain network maps. These correspond to the average of the model-based estimates of subject-specific brain networks across subjects in the study.  From this window, the user can select different ICs to display various brain network estimates. Figure \ref{fig:groupViewer} displays the group level map of the IC corresponding to default mode network. By default, HINT displays the intensity of the estimated spatial source signals in the population-level brain network. Alternatively, by selecting the ``View Z-score" option, the user can view the Z-scores of the spatial source signals which are derived from the raw intensity. An advantage of the Z-score is that it is unit-free and standardized, making it easier to compare across different ICs and to threshold. In fMRI analysis, researchers are often interested in thresholded IC maps to identify ``significantly activated" voxels in a brain network. HINT allows user to threshold the IC maps by either specifying a particular  Z-score value or by moving the Z thresholding slider until it reaches a satisfactory thresholded IC map. Then the user can save the thresholded IC map using the ``Create Mask" option. The saved threshold IC mask can be used in other display viewer windows to view the effects within an estimated brain network.

\subsubsection{Sub-population display maps}

The second display viewer, ``Sub-population display maps" shows the the sub-population brain network maps.  In this window, the user can define a sub-population of interest by specifying a combination  of covariate values $\boldsymbol{x}^{\ast}$. The window then displays the estimated sub-population IC maps as defined in equation (\ref{subpop}) for the corresponding sub-population. HINT also allows the user to specify multiple sub-populations of interest and display their estimated brain networks side by side in this display viewer.  This helps the user to visually compare the networks between sub-populations, such as treatment vs. control group or diseased vs normal group.  Figure \ref{fig:subpopCompare} in the main manuscript shows an example where we specify two sub-populations: subpop 1: people in treatment group with a score of 28 and gender 0 and subpop 2: people in the control group with a score of 45 and gender 0.  The left section of the display window provides the side-by-side viewing of the estimated maps of the default mode network for the two sub-populations.  As in the population display window, the user can use the Z thresholding slide bar to view the thresholded IC maps at different Z-value thresholds.  The crosshair movement is synced across the sub-population IC maps to facilitate the comparison of the spatial source signals between sub-populations across voxels in the network.

\subsubsection{Subject specific display maps}

\begin{figure}[H]
	\centering
	\includegraphics[width=1\linewidth]{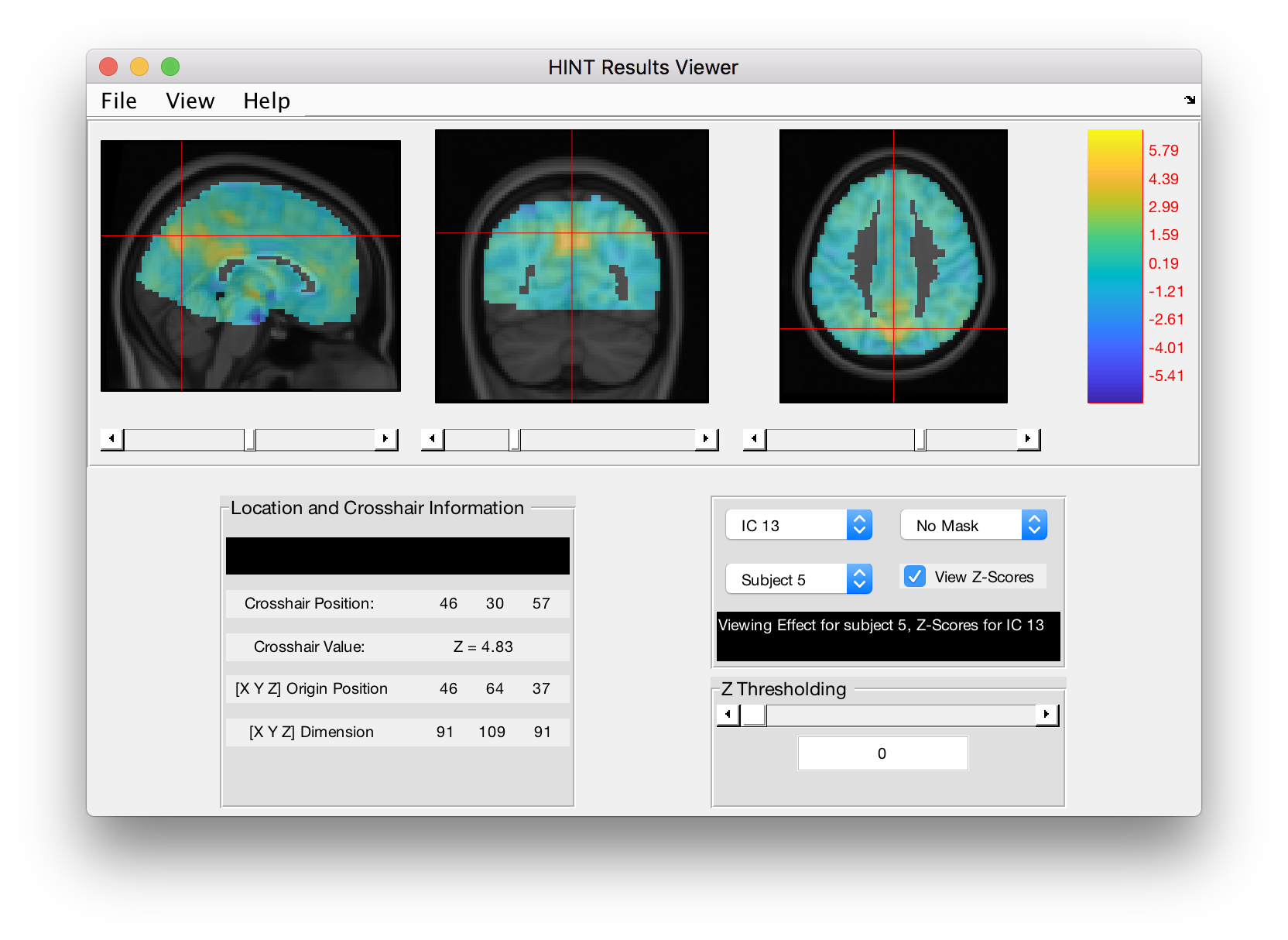}
	\caption{The HINT single-subject viewer. The user can select the estimated maps for each individual subject as well as apply masks generated using the population-level viewer.}
	\label{fig:singleSubjViewer}
\end{figure}

The third display viewer, ``Subject specific display maps", allows an investigator to view the estimated IC maps for each subject in the study. These maps can be viewed much in the same way as the population maps. It is also possible to apply the thresholed masks generated from the population-level display viewer to the subject-specific images to show the estimated networks across subjects.  See Figure \ref{fig:singleSubjViewer} for an example of this window.

\subsubsection{Beta coefficient display maps}

\begin{figure}[H]
	\centering
	\includegraphics[width=1\linewidth]{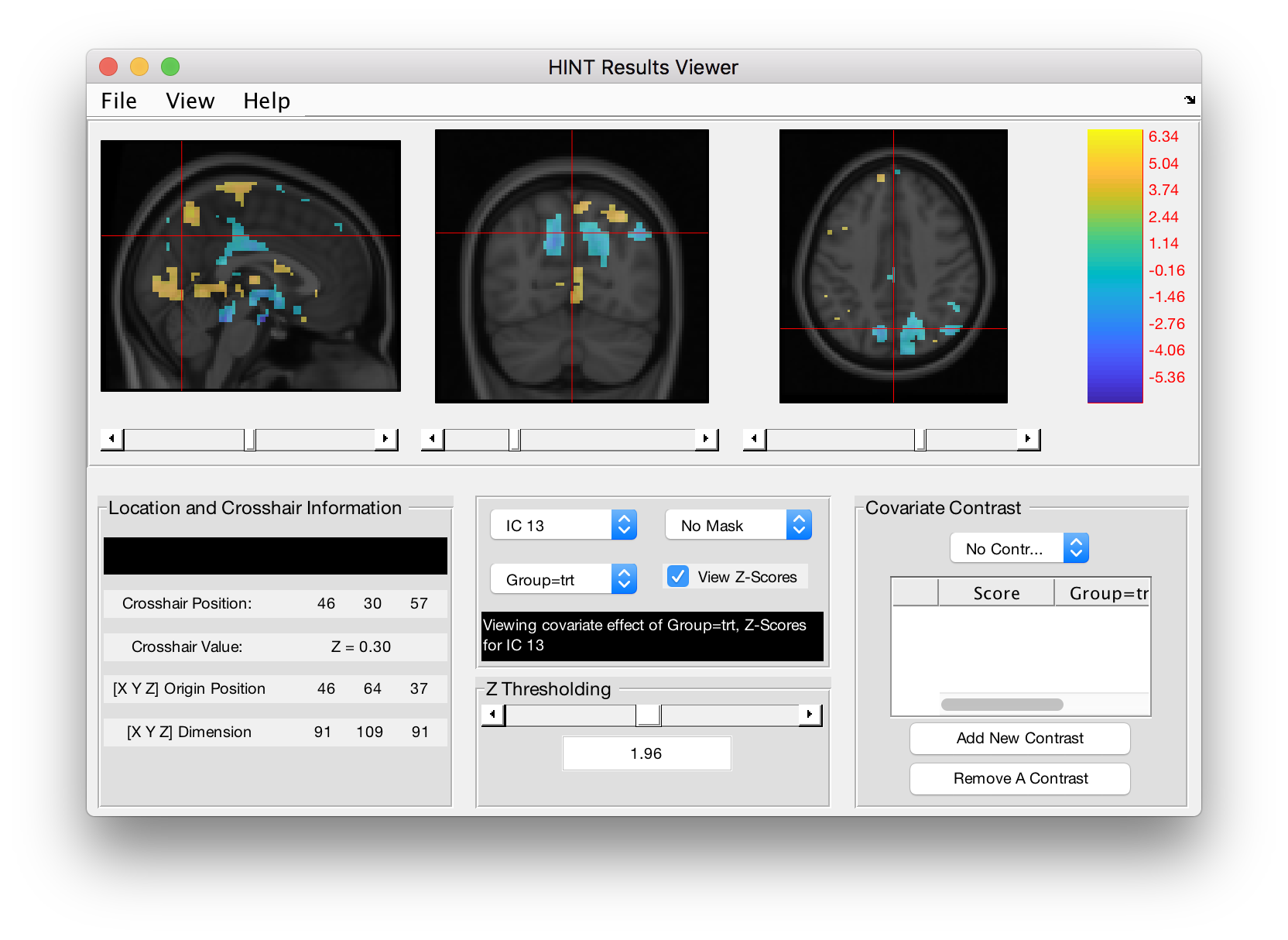}
	\caption{Example of the covariate effect viewer for the effect of being in the treatment group on the IC corresponding to the default mode network.}
	\label{fig:covEffectViewer}
\end{figure}

\begin{figure}[H]
	\centering
	\includegraphics[width=1\linewidth]{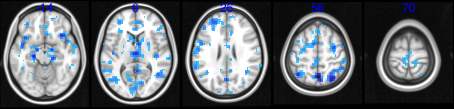}
	\caption{ The estimated treatment effect in the DMN under TC-GICA  using a $Z = 1.96$ significance threshold (compare to Figure \ref{fig:covEffectViewer}). }
	\label{fig:DRcompare}
\end{figure}

The fourth display viewer, Beta coefficient display maps (Figure \ref{fig:covEffectViewer}), shows the estimated maps of the beta coefficients and their linear combinations. These displays allow users to view the covariate effects within the estimated brain networks and identify locations in the networks that show significant covariate effects. Hypothesis tests can be performed using the inference procedure described in Section 2.4 and test results can be shown in the display viewer. The resulting maps are shown by Z-score, and can be thresholded using the slider bar or by specifying an exact threshold. As an example, Figure \ref{fig:DRcompare} shows the Z-score thresholded at 1.96 for testing the treatment effects in the example data set. For comparison purpose, Figure \ref{fig:DRcompare} also displays the corresponding Z-score map for testing treatment effects based on TC-GICA. The effect estimates using TC-GICA are more spatially dispersed than those estimated using hc-ICA, for example the second slice in Figure \ref{fig:DRcompare} shows a large area of activation in the frontal region of the brain. These spatially dispersed estimates are indicative of noiser estimates from TC-GICA. On the other hand, HINT shows more significant treatment effects within the relevant regions in the network. Thus, by formally modeling covariate effects in the ICA decomposition, HINT provides more accurate estimates and demonstrates higher statistical power in detecting covariate effects than TC-GICA.

{\it Contrast Specification}

\begin{figure}[H]
	\centering
	\includegraphics[width=1\linewidth]{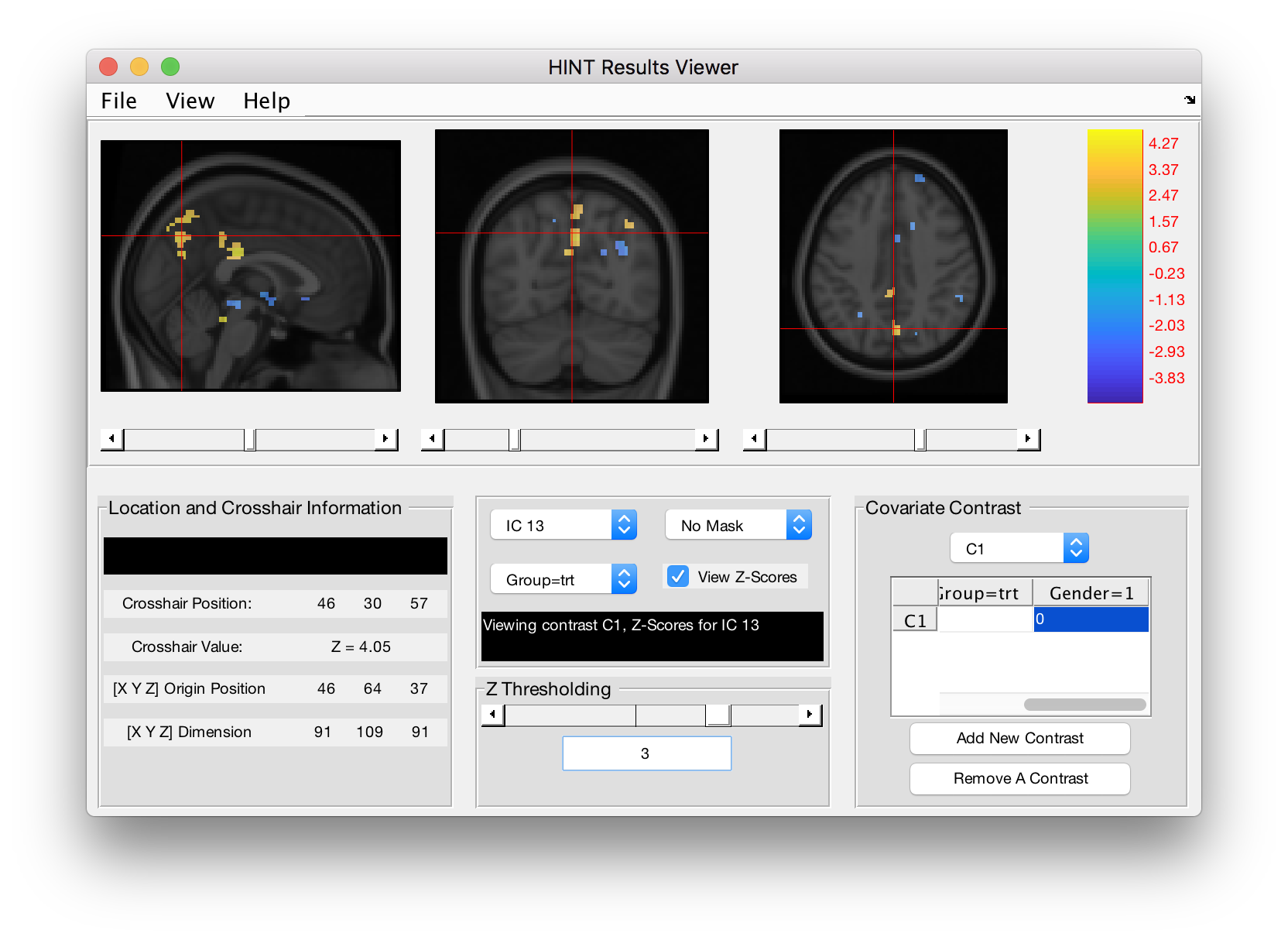}
	\caption{Example of the contrast for people in the treatment group with a score of 30 and of the gender coded as 0. The image is thresholded using $Z=3$ as the cutoff.}
	\label{fig:contrastExample}
\end{figure}

The bottom-right panel in Figure \ref{fig:covEffectViewer} is the contrast specification panel for the covariate effects. Say we are interested in the overall effect on the brain network for a person with a score of 30 in the treatment group who was also of the Gender coded as 0. Recall that our covariate effects, in order, are (1) score, (2) being in the treatment group, and (3) being gender 1. The corresponding vector of contrast coefficients is $\bm \lambda' = [30 \; 1 \; 0]$. In the bottom-right panel of Figure \ref{fig:covEffectViewer}, we select ``add new contrast", and fill out the values for the main effects. Any specified interactions are automatically calculated based on the provided values. Then, selecting the contrast from the drop-down menu above the contrast list, we see the image displayed in Figure \ref{fig:contrastExample}. This is the contrast image for testing hypotheses about the specified linear combination of covariate effects. Notice that this map can be thresholded by Z-score.

\newpage
\section*{Appendix 2 - Runinfo File Structure}

\begin{table}[H]
\caption{The objects contained in the runinfo file.}
\begin{center}
\begin{tabular}{|l|c|}
	\hline
	Variable & Description \\
	\hline
	N & The number of subjects \\
	X & The design matrix \\
	varNamesX & The names of the columns of the design matrix \\
	varInModel & Whether a variable is included in the model\\
	interactions & The interactions in the model in terms of the design matrix \\
	interactionsBase & The interactions in the model in terms of the original covariates \\
	YtildeStar & The $Nq \times V$ matrix of preprocessed data \\
	beta0Star & The initial guess for the beta maps \\
	covariates & The covariate names \\
	covfile & The filepath to the covariate file \\
	isCat & A $p \times 1$ vector indexing categorical covariates \\
	maskfl & The path to the mask file \\
	niifiles & A cell array of paths to the subject level fMRI data \\
	numPCA & The number of principal components for preprocessing \\
	outfolder & Path to the output directory \\
	prefix & Prefix for the files in the analysis \\
	q & The number of independent components \\
	thetaStar & Structure containing initial guess values \\
	time\_num & The number of time points for each subject \\
	voxSize & The dimension of the mask \\
	\hline
\end{tabular}
\end{center}
\label{runinfoTable}
\end{table}

Table \ref{runinfoTable} displays the variables contained in the runinfo file for a HINT analysis.

%
%
\newpage
\section*{Appendix 3 - Full Synthetic Data Results}

\begin{table}[H]
\setlength\tabcolsep{-8.5pt}
\centering
\hspace*{-0.2cm}\begin{tabular}{ccc}
\multicolumn{3}{c}{No Overlap} \\
\includegraphics[scale=.3]{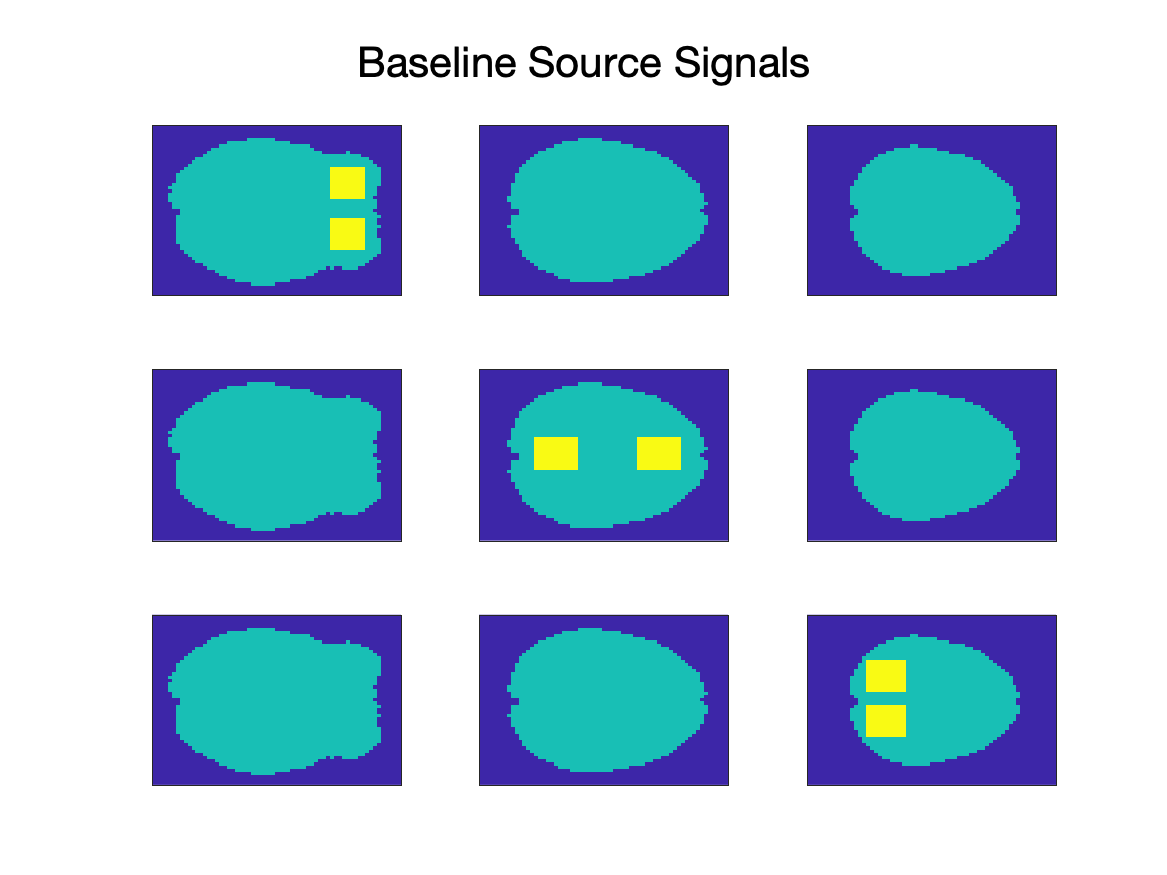} & 
\includegraphics[scale=.3]{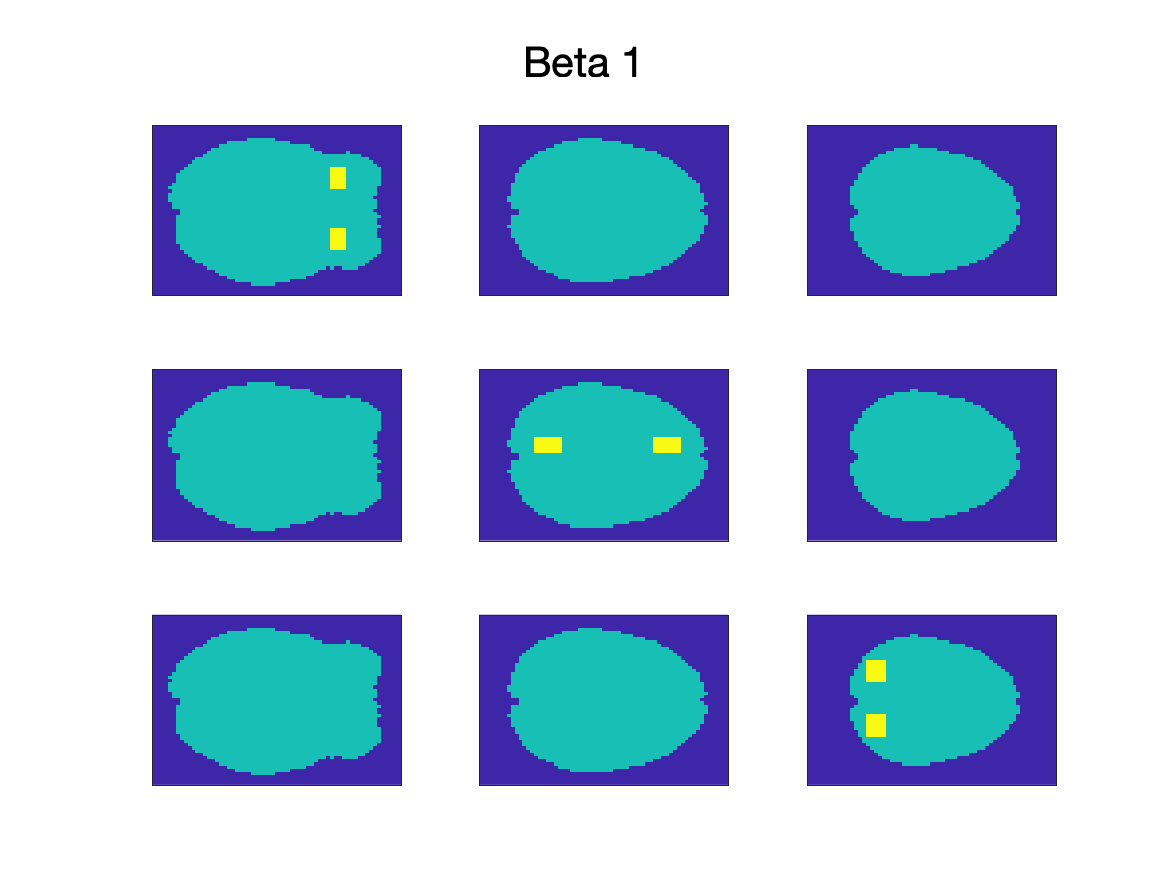} & 
\includegraphics[scale=.3]{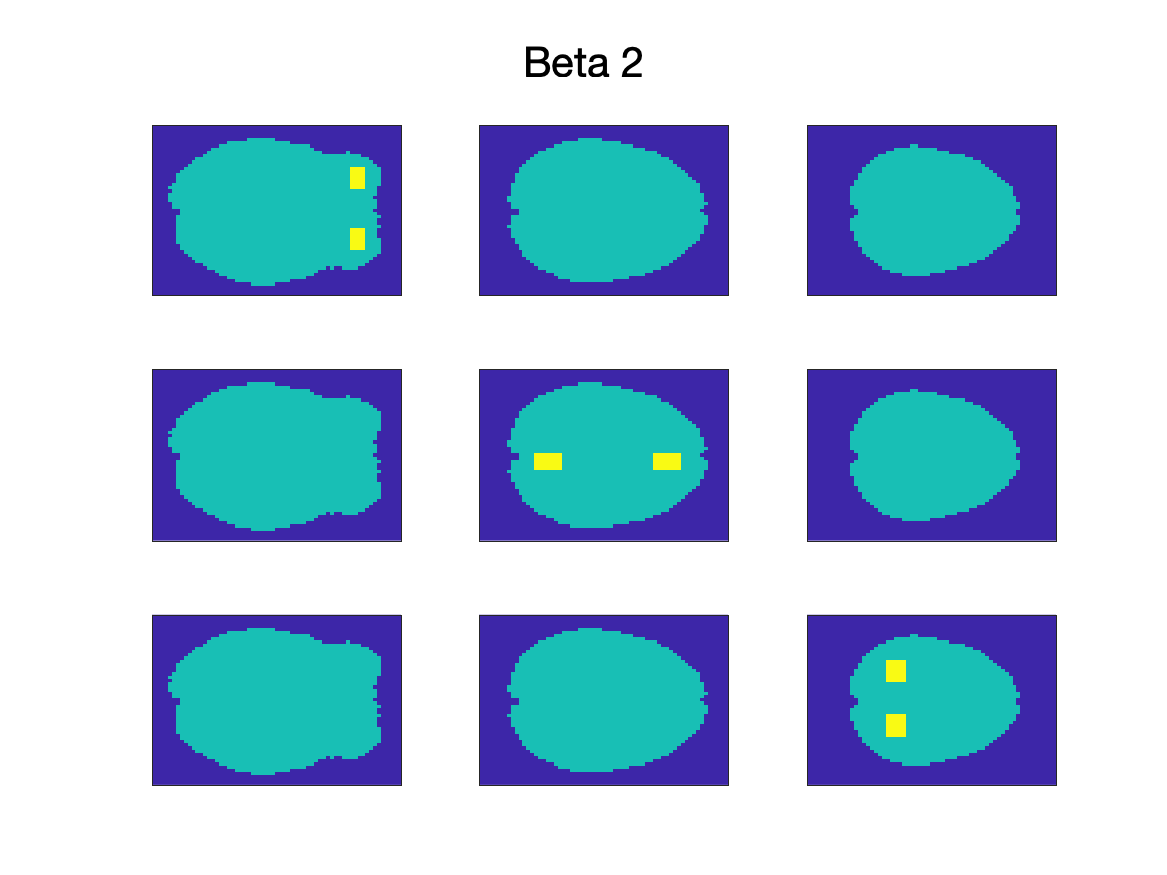} \\

\multicolumn{3}{c}{Beta Overlap} \\
\includegraphics[scale=.3]{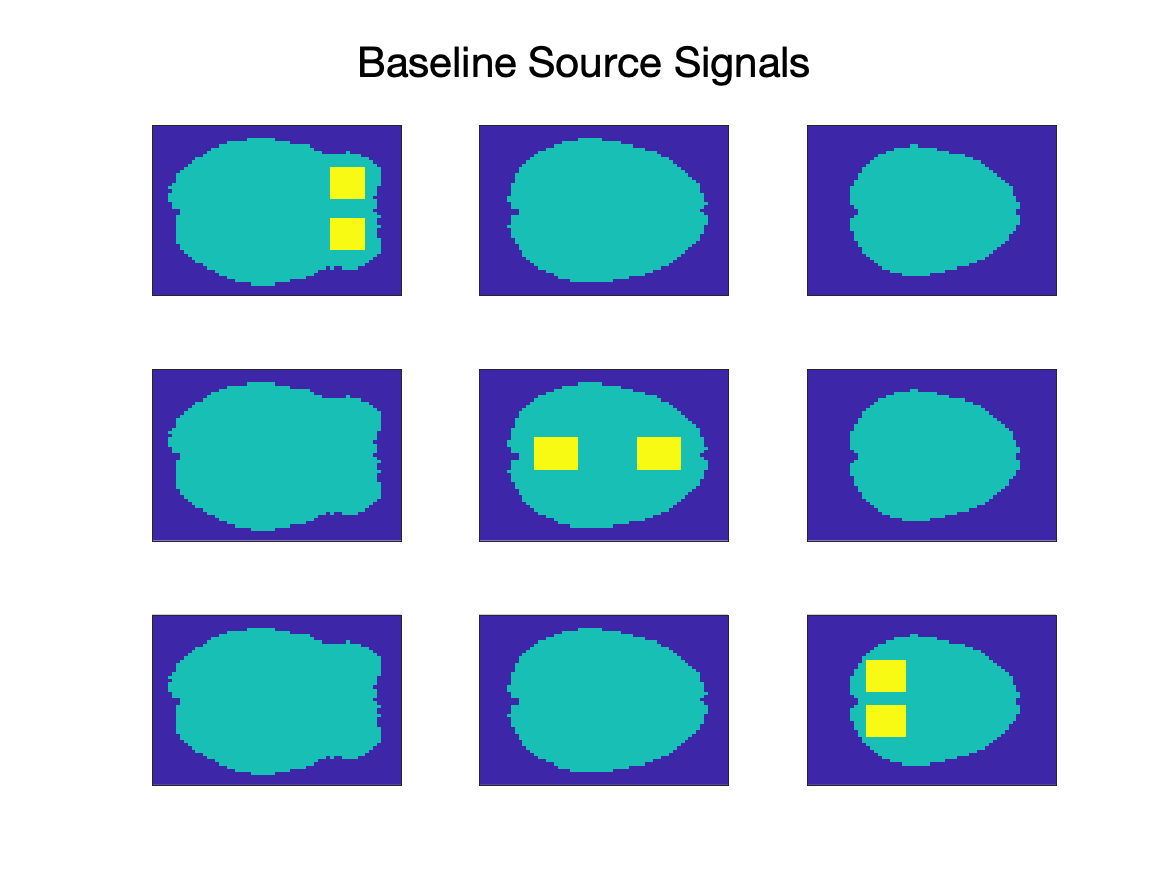} & 
\includegraphics[scale=.3]{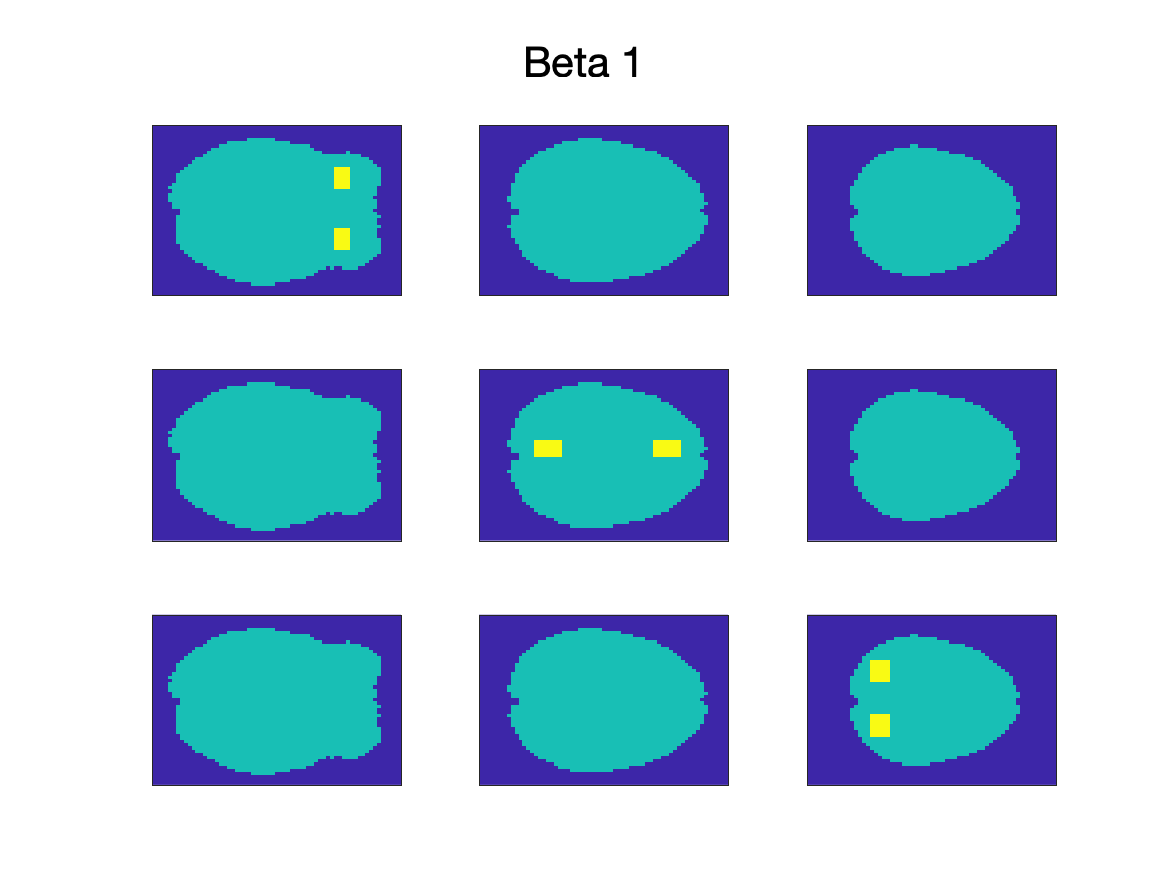} & 
\includegraphics[scale=.3]{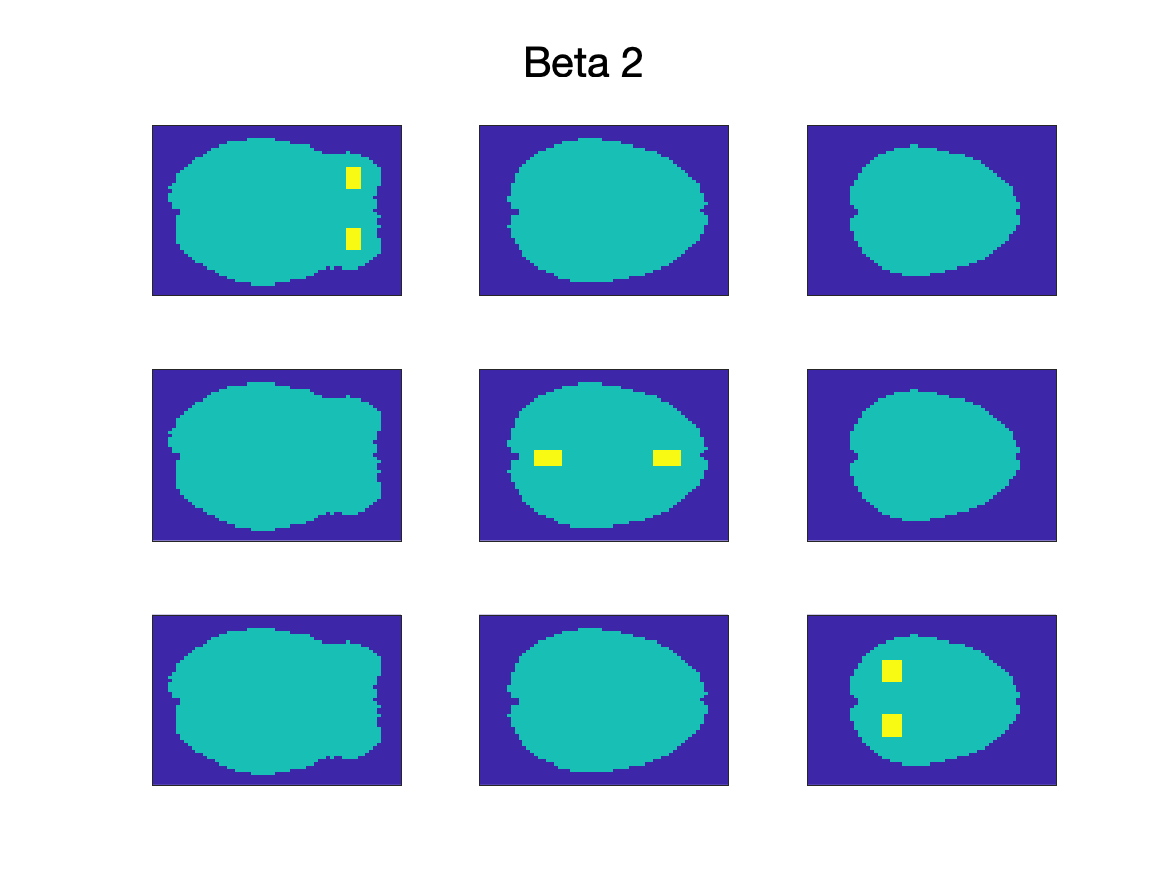} \\

\multicolumn{3}{c}{S Overlap} \\
\includegraphics[scale=.3]{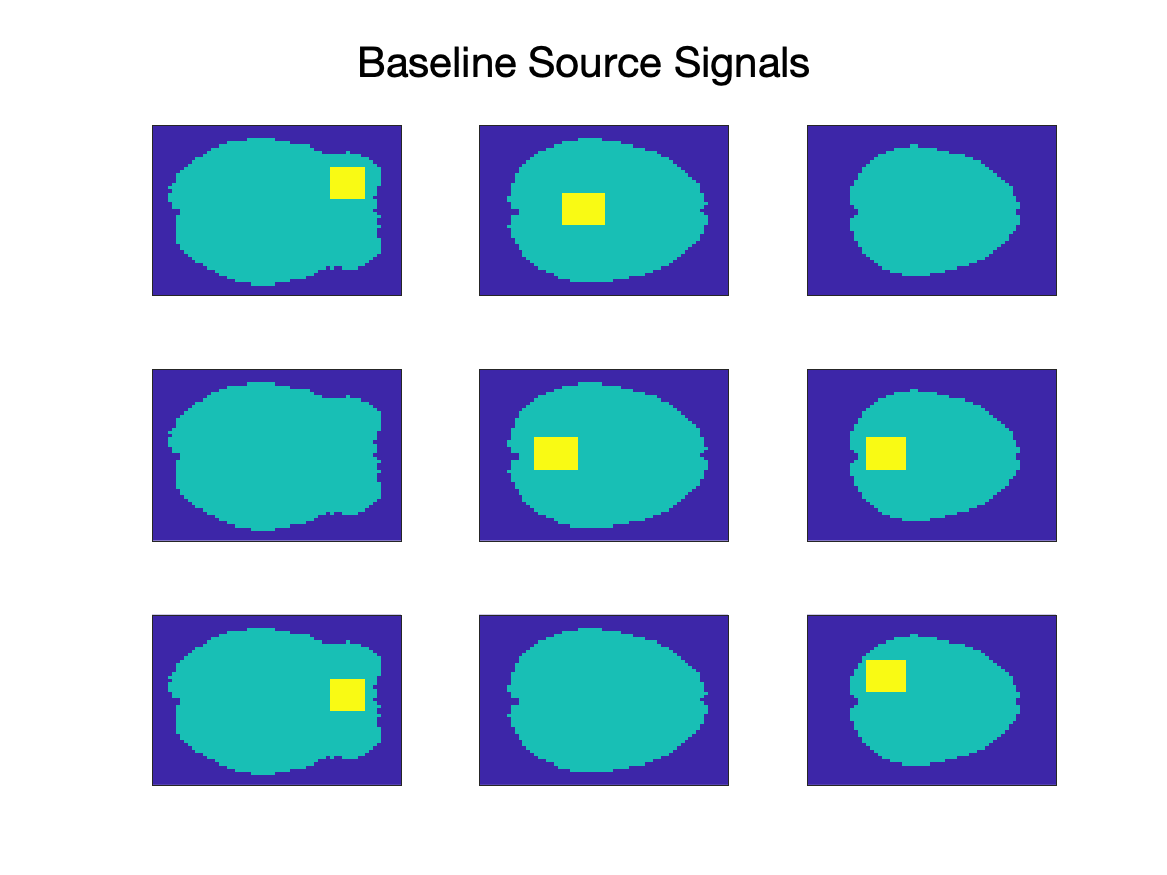} & 
\includegraphics[scale=.3]{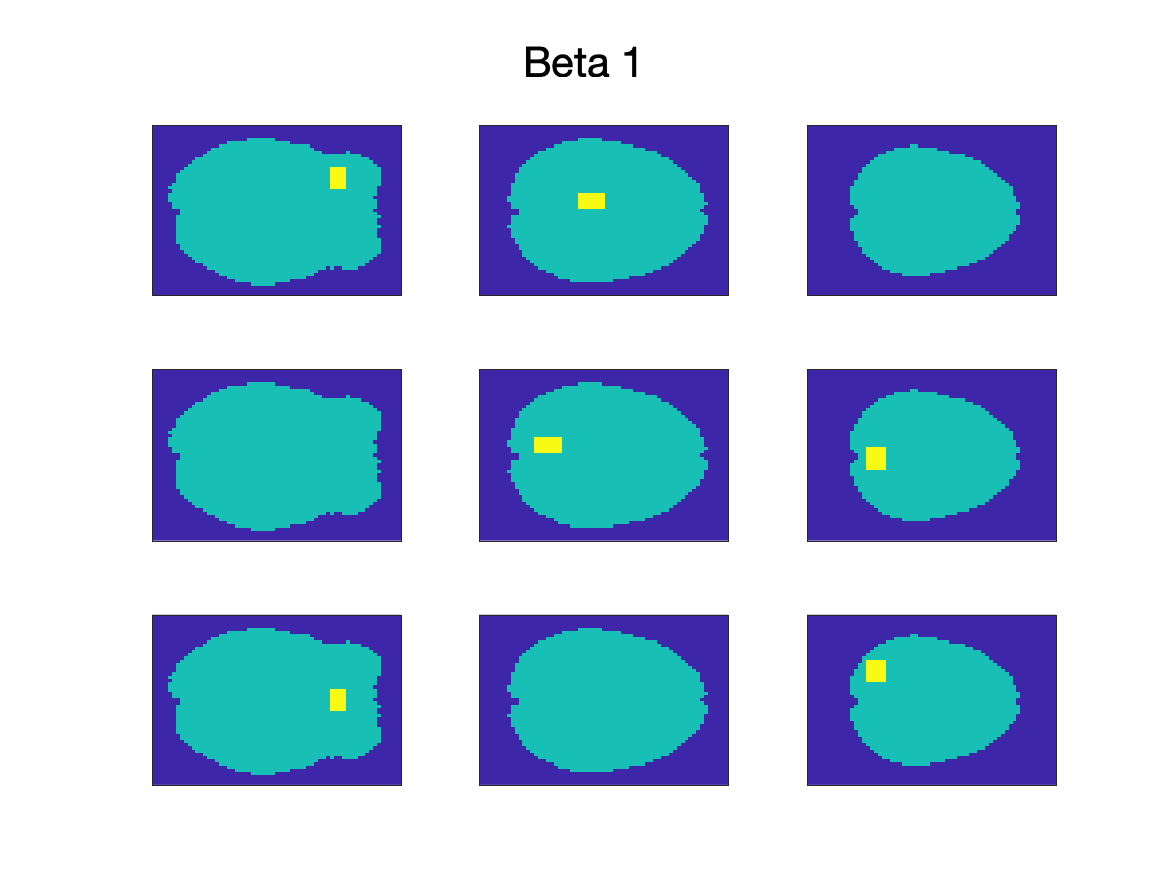} & 
\includegraphics[scale=.3]{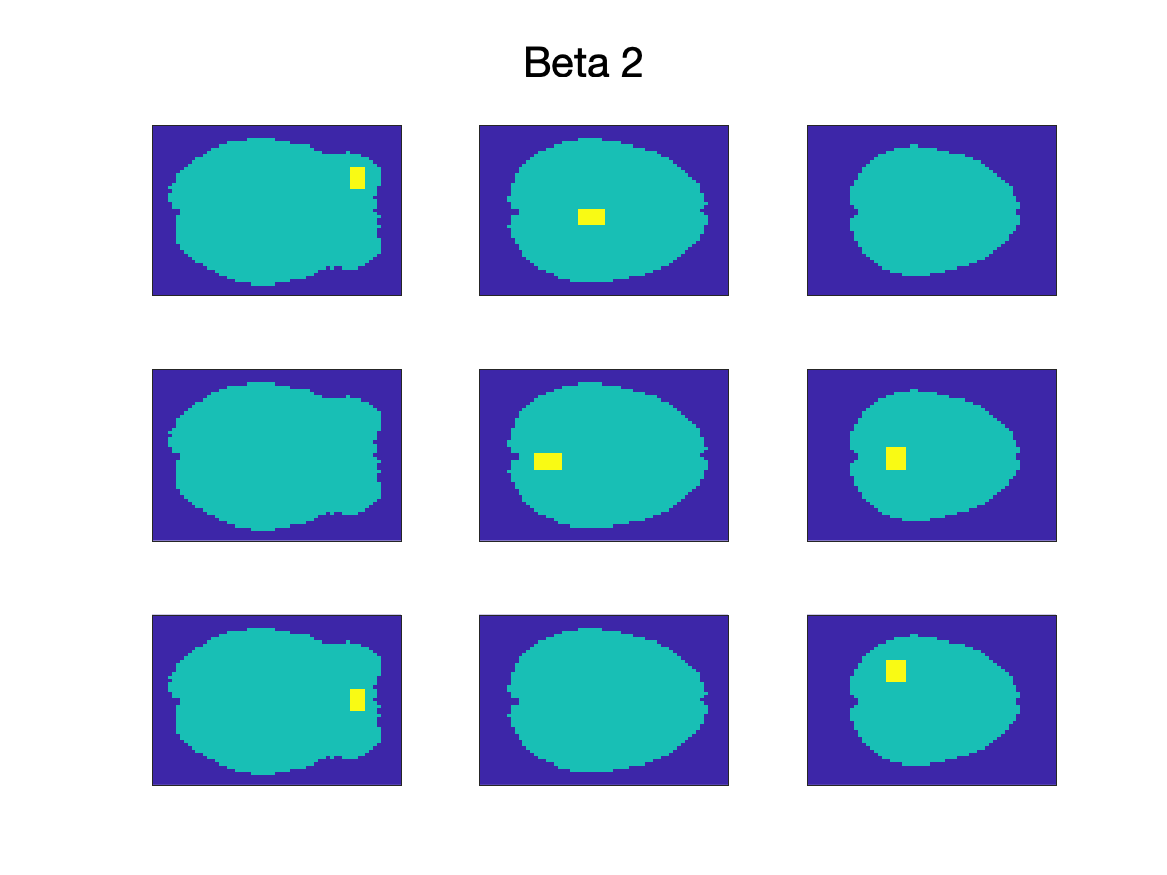} \\

\multicolumn{3}{c}{S and Beta Overlap} \\
\includegraphics[scale=.3]{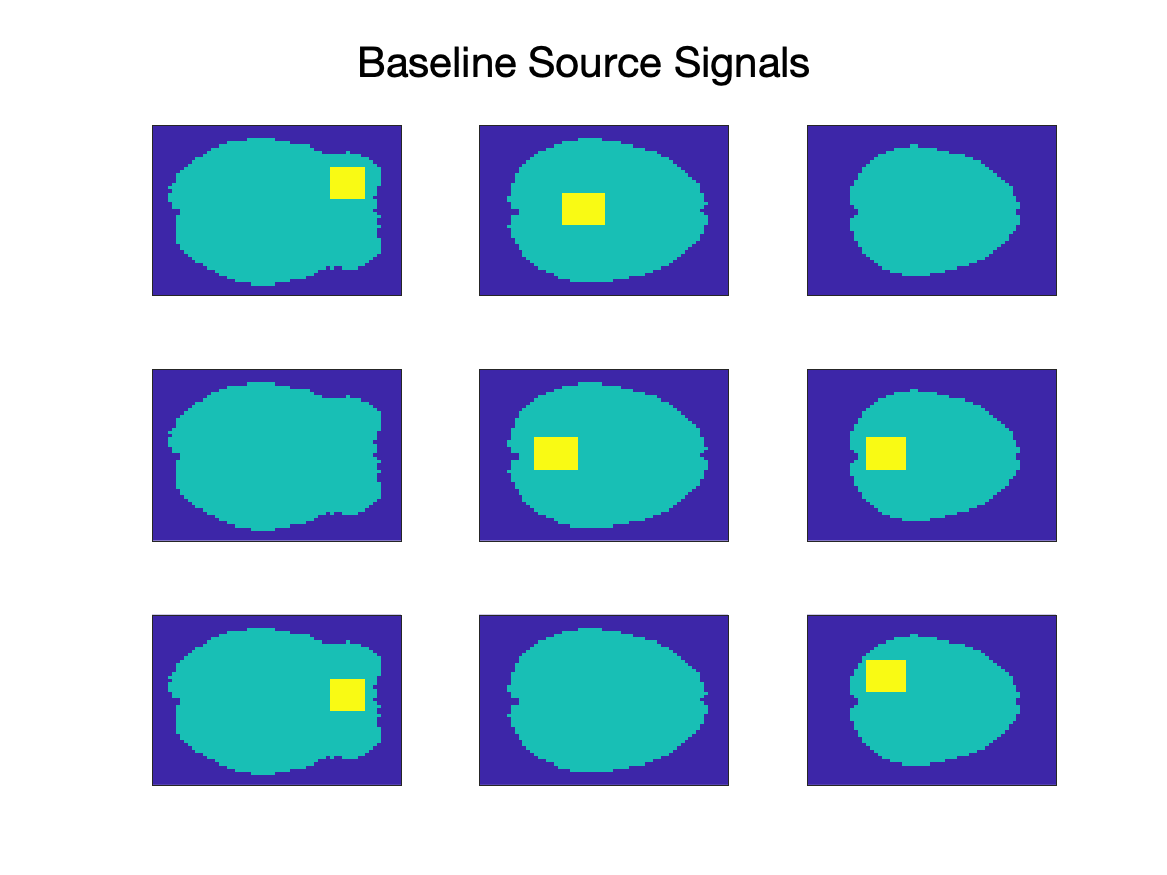} & 
\includegraphics[scale=.3]{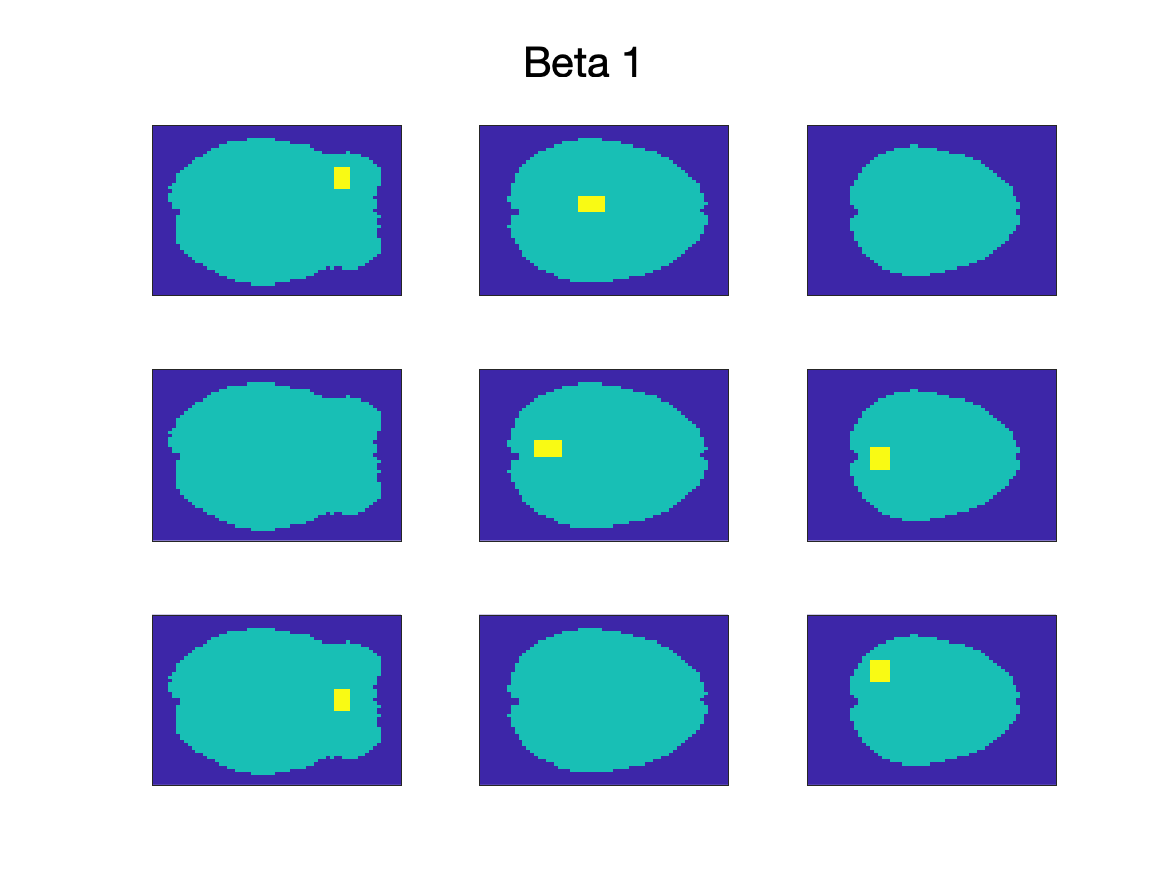} & 
\includegraphics[scale=.3]{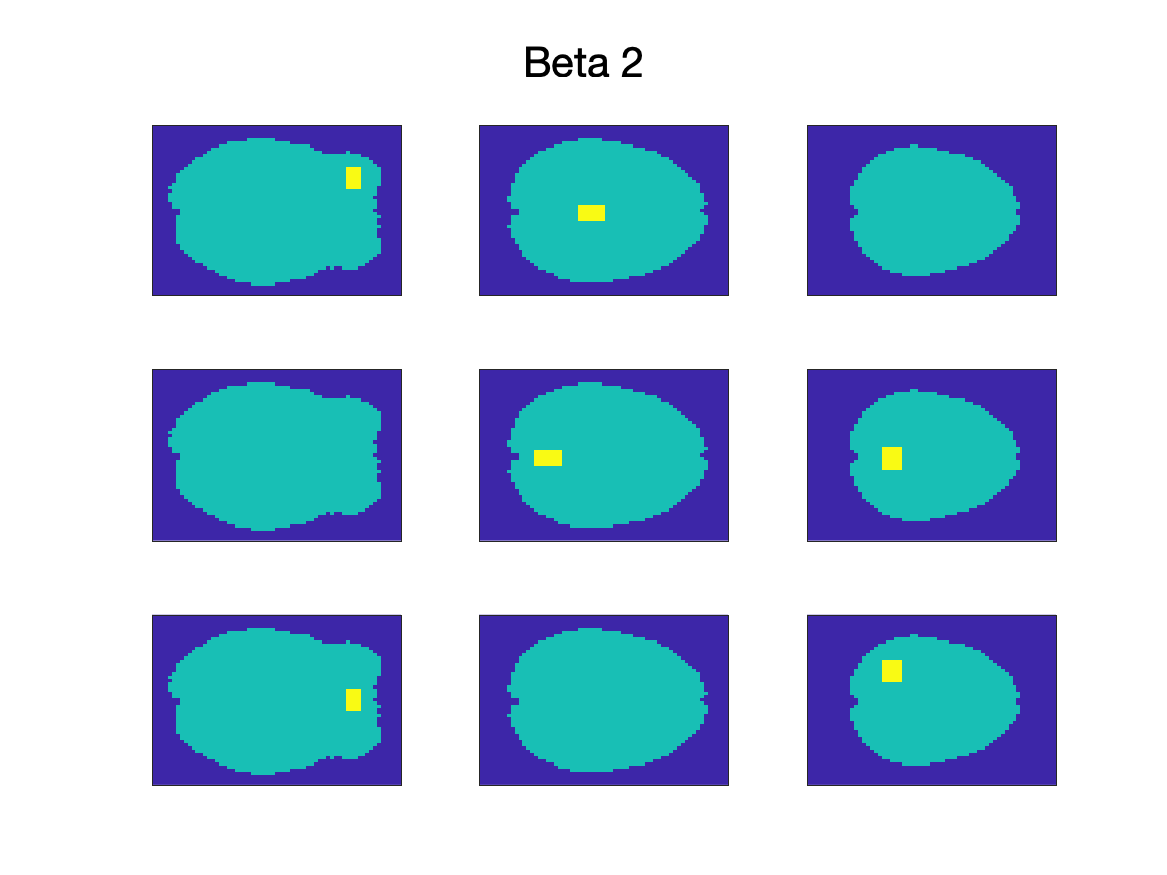} \\
\end{tabular}
\caption{Spatial maps used for the synthetic data study. Within each image, each row corresponds to an IC. The first set of maps correspond to the baseline source signals. The middle set of maps corresponds to the covariate effect for the first covariate. The final set of maps corresponds to the covariate effect for the second covariate.}
\label{fig:allmaps}
\end{table}

\begin{figure}[H]
	\centering
	\includegraphics[width=0.95\linewidth]{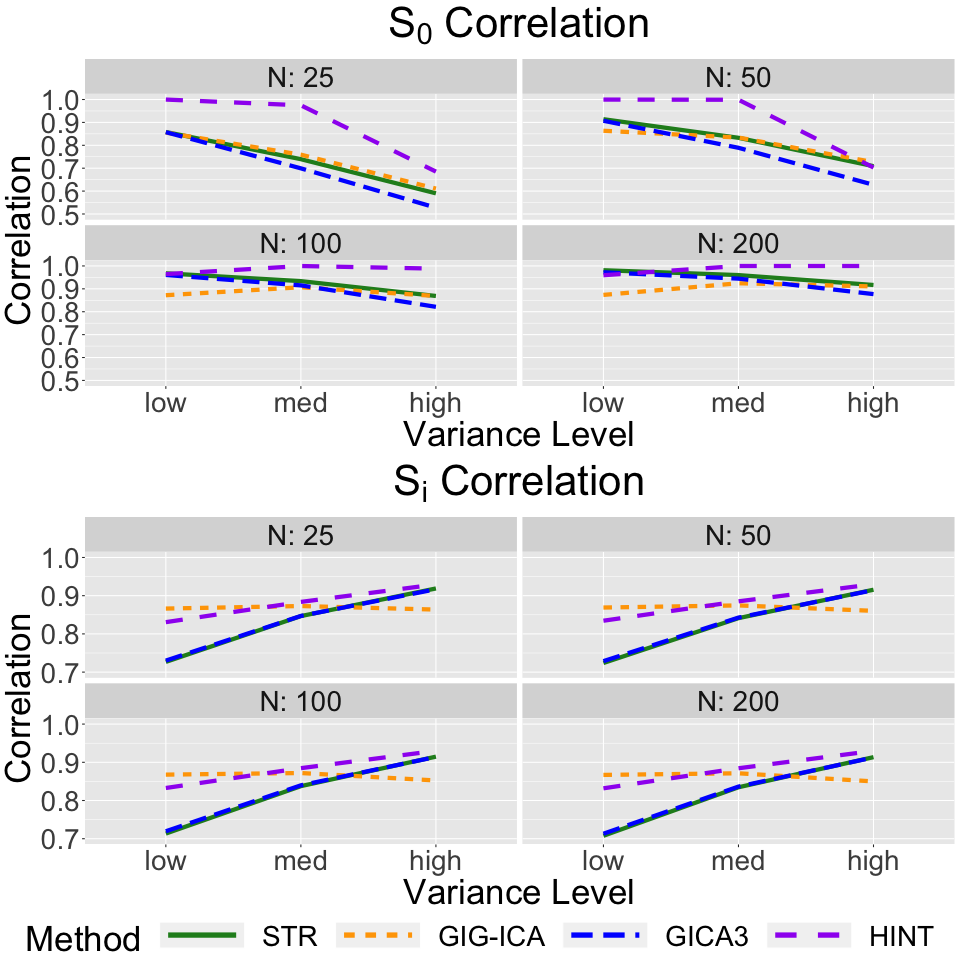}
	\caption{Correlation with the true spatial maps under STR, GICA3, GIG-ICA, and HINT for the synthetic data with no spatial overlap in the baseline source signals but some spatial overlap in the covariate effect maps.}
	\label{fig:so0bo1_cor}
\end{figure}

\begin{figure}[H]
	\centering
	\includegraphics[width=0.95\linewidth]{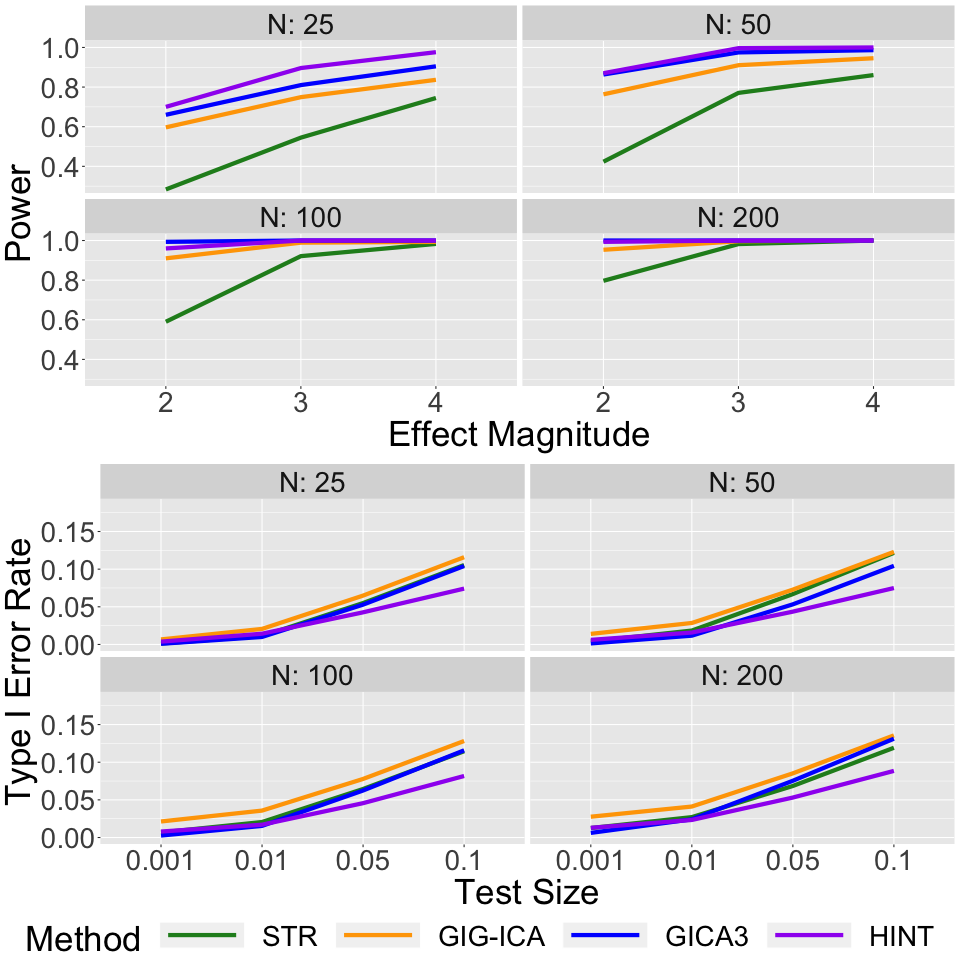}
	\caption{Power and Type I Error Rates under STR, GICA3, GIG-ICA, and HINT for the synthetic data with no spatial overlap in baseline source signals but some spatial overlap in the covariate effects.}
	\label{fig:so0bo1_powert1}
\end{figure}

\begin{figure}[H]
	\centering
	\includegraphics[width=0.95\linewidth]{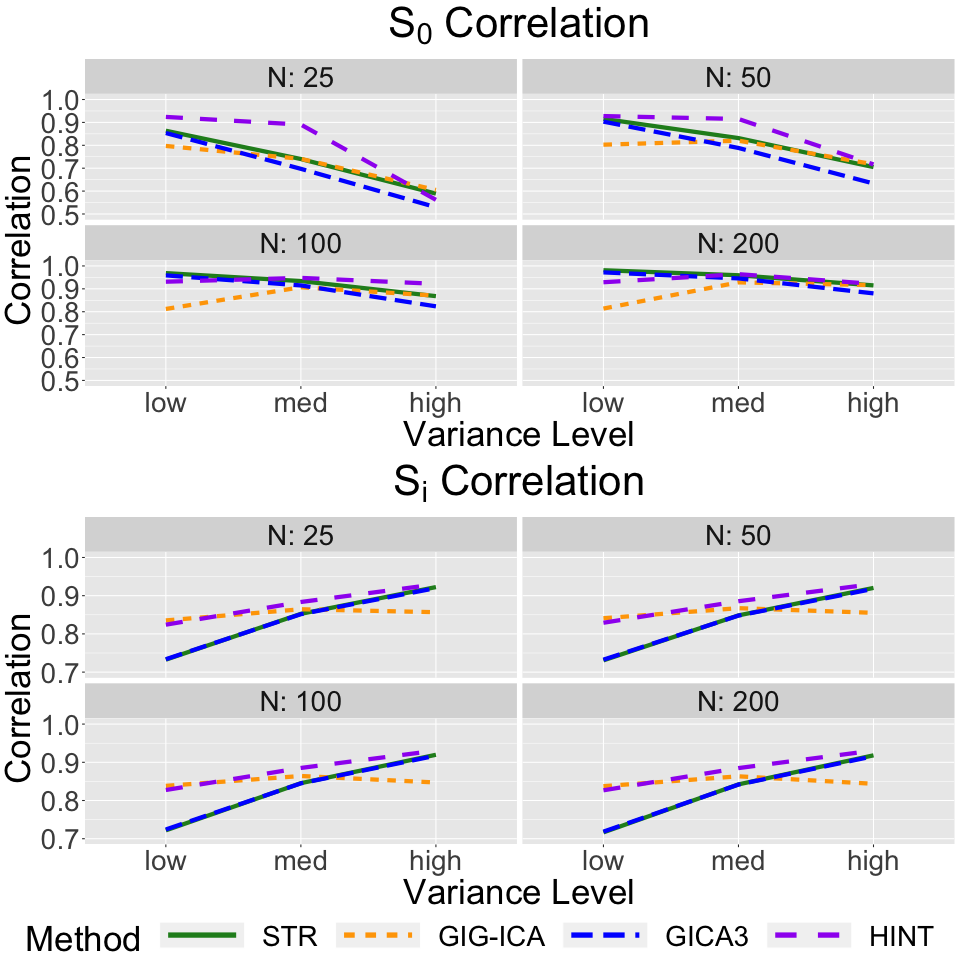}
	\caption{Correlation with the true spatial maps under STR, GICA3, GIG-ICA, and HINT for the synthetic data with spatial overlap in the baseline source signals but no spatial overlap in the covariate effect maps.}
	\label{fig:so1bo0_cor}
\end{figure}

\begin{figure}[H]
	\centering
	\includegraphics[width=0.95\linewidth]{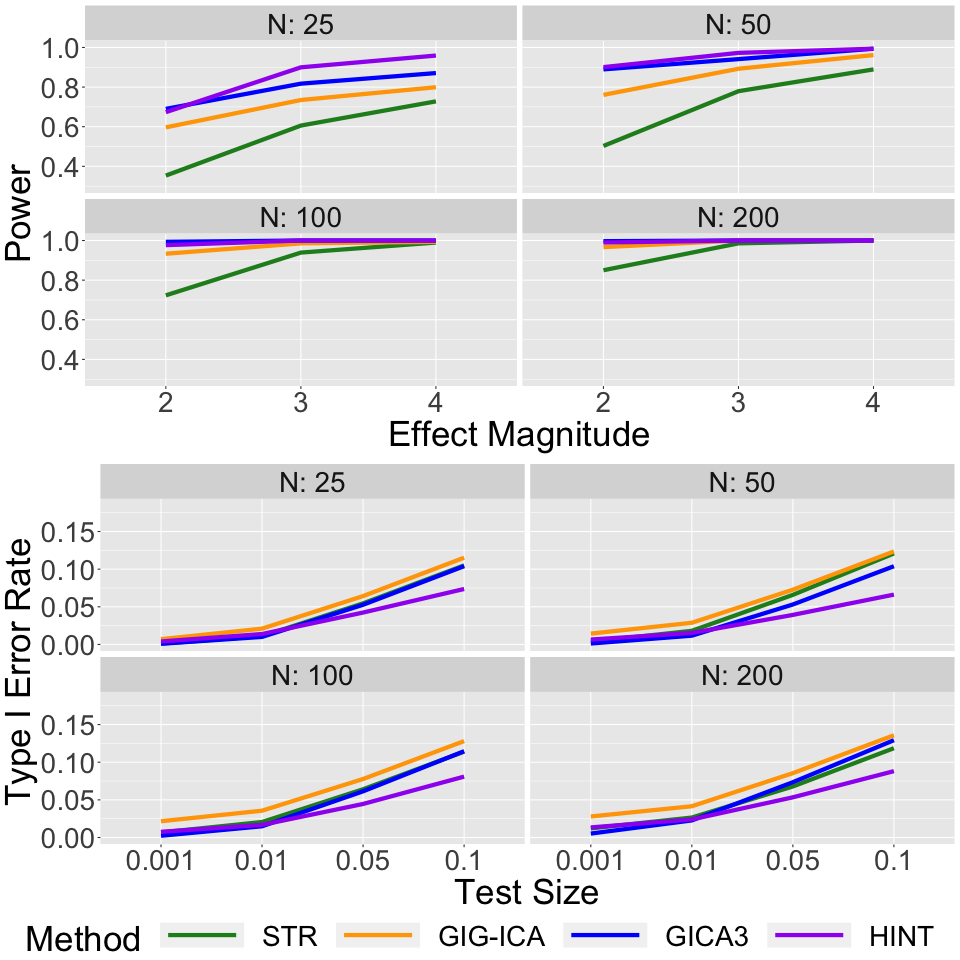}
	\caption{Power and Type I Error Rates under STR, GICA3, GIG-ICA, and HINT for the synthetic data with spatial overlap in baseline source signals but no spatial overlap in the covariate effects.}
	\label{fig:so1bo0_powert1}
\end{figure}

\begin{figure}[H]
	\centering
	\includegraphics[width=0.95\linewidth]{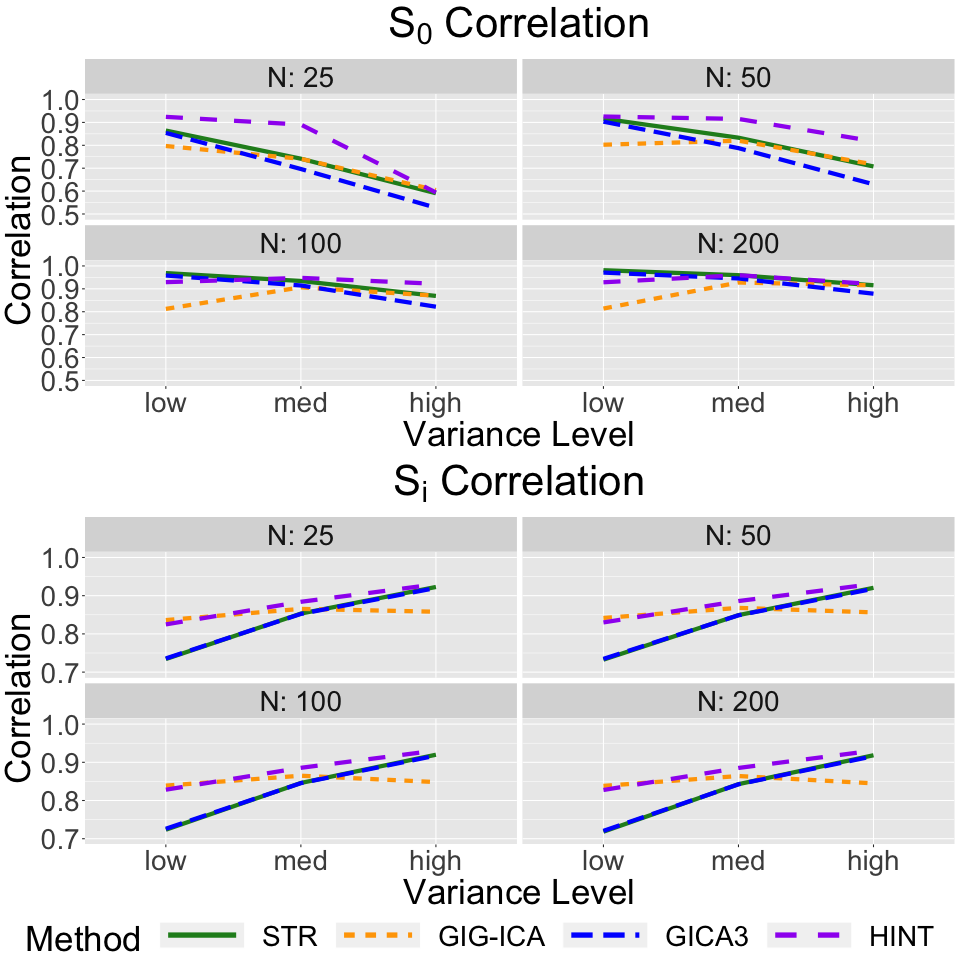}
	\caption{Correlation with the true spatial maps under STR, GICA3, GIG-ICA, and HINT for the synthetic data with spatial overlap in both the baseline source signals and the covariate effect maps.}
	\label{fig:so1bo1_cor}
\end{figure}

\begin{figure}[H]
	\centering
	\includegraphics[width=0.95\linewidth]{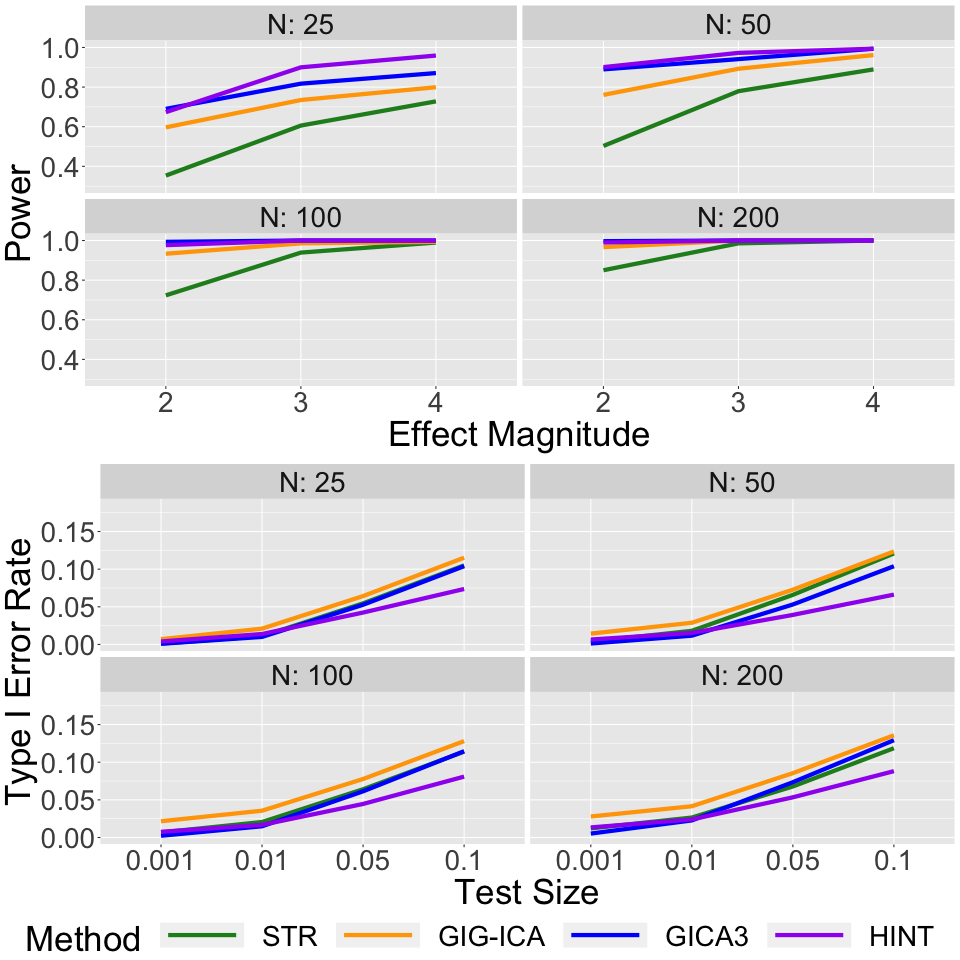}
	\caption{Power and Type I Error Rates under STR, GICA3, GIG-ICA, and HINT for the synthetic data with spatial overlap in both the baseline source signals and the covariate effects.}
	\label{fig:so1bo1_powert1}
\end{figure}

\end{document}